\documentclass[twocolumn,prl,aps,epsf]{revtex4}
\usepackage{graphicx}

\begin{document}
\title{One- and many-body effects on mirages in quantum corrals}
\author{A. Lobos and A. A. Aligia}
\address{Centro At\'{o}mico Bariloche and Instituto Balseiro, Comisi\'{o}n Nacional\\
de Energ\'{\i }a At\'{o}mica, 8400 Bariloche, Argentina.}
\begin{abstract}

Recent interesting experiments used scanning tunneling microscopy to study
systems involving Kondo impurities in quantum corrals assembled on Cu or
noble metal surfaces. The solution of the two-dimensional one-particle
Schr\"{o}dinger equation in a hard wall corral without impurity is useful to
predict the conditions under which the Kondo effect can be projected to a
remote location (the quantum mirage). To model a soft circular corral, we
solve this equation under the potential $W\delta (r-r_{0})$, where $r$ is
the distance to the center of the corral and $r_{0}$ its radius. We expand
the Green's function of electron surface states $G_{s}^{0}$ for $r<r_{0}$ as
a discrete sum of contributions from single poles at energies $\epsilon
_{i}-i\delta _{i}$. The imaginary part $\delta _{i}$ is the half-width of
the resonance produced by the soft confining potential, and turns out to be
a simple increasing function of $\epsilon _{i}$. In presence of an impurity,
we solve the Anderson model at arbitrary temperatures using the resulting
expression for $G_{s}^{0}$ and perturbation theory up to second order in the
Coulomb repulsion $U$. We calculate the resulting change in the differential
conductance $\Delta dI/dV$ as a function of voltage and space, in circular
and elliptical corrals, for different conditions, including those
corresponding to recent experiments. The main features are reproduced. The
role of the direct hybridization between impurity and bulk, the confinement
potential, the size of the corral and temperature on the intensity of the
mirage are analyzed. We also calculate spin-spin correlation functions.
\end{abstract}


\maketitle
\narrowtext

\section{Introduction}
\label{Introduction} In recent years, the advances in scanning tunneling
microscopy (STM) made possible the manipulation of single atoms on top of a
surface and the construction of quantum structures of arbitrary shape \cite
{eig}. In particular, quantum corrals have been assembled by depositing a
close line of atoms or molecules on Cu or noble metal (111) surfaces \cite
{cro,hel,man,man2}. These surfaces have the property that for small wave
vectors parallel to the surface a parabolic band of two-dimensional (2D)
surface states uncoupled to bulk states exists \cite{hub}. A circular corral
of radius $71.3$ \AA\ was constructed depositing 48 Fe atoms on a Cu(111)
surface \cite{cro}. The measured differential conductance $dI/dV$ could be
fitted by a combination of the density of eigenstates close to the Fermi
energy $\epsilon _{F}$, of a 2D electron gas inside a hard wall circular
corral \cite{cro}. On another experiment, a Co atom acting as a magnetic
impurity was placed at one focus of an elliptical quantum corral, and the
corresponding Kondo feature in $dI/dV$ was observed not only at that
position, but also with reduced intensity at the other focus \cite{man}.
This {\it ``}mirage{\it ''} can be understood as the result of quantum
interference in the way in which the Kondo effect is transmitted from one
focus to the other by the different conduction eigenstates of a hard-wall
ellipse \cite{rap}. Among these eigenstates, the density of the one closest
to $\epsilon _{F}$ is clearly displayed in the change of differential
conductance $\Delta dI/dV$ after adding the impurity \cite{man}. More
recently, the extrema in the degenerate $37^{th}$ and $38^{th}$ conduction
eigenstates of a hard wall circular corral have inspired an experiment with
two impurities inside the corral \cite{man2}.

The theories of the mirage experiment can be classified into those which
start from eigenstates of a confined 2D electron gas \cite
{rap,por,wei,hal,ali2,wil} and those in which the confining atoms are
modeled by a phenomenological scattering approach \cite{aga,fie}. The latter
have the disadvantage that many-body effects are very hard to include. In
addition, some features which are clear from the hard wall eigenstates, like
mirages out of the foci \cite{man2,rap,por,wei}, are somewhat hidden in the
scattering approaches. On the other hand, while the eigenstates inside a
hard wall corral are perfectly defined, in the actual experiments the
boundaries of the corrals are soft and the eigenstates become resonances
with finite width $\delta _{i}$ \cite{cro}. This width plays a crucial role
in the line shape of $\Delta dI/dV$ and its magnitude at the mirage point 
\cite{rap,ali2}. If $\delta _{i}$ is large the mirage disappears, while if $%
\delta _{i}=0$ and other parameters as in the experiment, there is no Kondo
resonance at $\epsilon _{F}$. Ordinary Lanczos calculations have $\delta
_{i}=0$ \cite{hal} and are unable to describe the line shape of $\Delta
dI/dV $. This can be corrected using an embedding method \cite{wil}. To the
best of our knowledge, this theory and the perturbative one \cite{rap,ali2}
using a constant value for $\delta _{i}=\delta $ as a parameter, are the
only many-body ones presented so far, which can explain the voltage
dependence of $\Delta dI/dV$. Thus, a careful study of the broadening $%
\delta _{i}$ is necessary.

Another important parameter which is still not well known is $\delta
_{b}/\delta _{s}$, where $\delta _{b}$ ($\delta _{s}$) is the part of the
resonant level width of the impurity state, which is due to hybridization
with bulk (surface) states. Some theories have assumed that $\delta _{b}\sim
\delta _{s}$, while others have taken $\delta _{b}=0$ \cite
{rap,por,wei,hal,ali2}. Clearly, $\delta _{s}\neq 0$, because otherwise $%
\Delta dI/dV$ would be independent of the 2D conduction eigenstates in
contrast to the experiment \cite{man}. Instead, the line shape for both, the
elliptical corral and the clean surface can be explained taking $\delta
_{s}=0$ and the same set of parameters \cite{ali2}. A recent experiment
suggests that $\delta _{b}>\delta _{s}$ on the basis of the rapid decay in $%
\Delta dI/dV$ as the STM tip is moved away from the impurity \cite{knorr}.
However, the 4s and 4p states of the impurity atom were not included in the
analysis and they can affect the distance dependence of $\Delta dI/dV$.
Recently, it has been suggested that experiments with two impurities inside
a quantum corral can elucidate the magnitude of $\delta _{b}/\delta _{s}$
because the effect of interactions between impurities is expected to be
proportional to $\delta _{s}^{2}$\cite{wil}.

In this work we solve the impurity Anderson model which describes the mirage
experiment using perturbation theory in the Coulomb repulsion $U$. The
previous theory of one of us \cite{rap} is extended to include non-vanishing 
$\delta _{b}$ and a more realistic $\delta _{i}$. We calculate the widths of
the conduction states $\delta _{i}$ for a circular corral. We obtain an
integral expression for the conduction electron Green's function $G_{s}^{0}$
in the absence of the impurity, and then we expand it as a discrete sum of
contributions from simple poles, using ideas borrowed from nuclear physics 
\cite{cal} and scattering theory \cite{tay}. This expansion takes a suitable
form for the perturbative approach. We also discuss the space dependence of
the compensation of the impurity spin. The calculation of $G_{s}^{0}$ is
presented in section II. In section III we describe the model and the
many-body approach. Section IV contains the results for $\Delta dI/dV$ in
circular and elliptical corrals with a magnetic impurity. In Section V, we
calculate the correlation functions between the impurity spin and that of
the conduction electrons as a function of position. Section VI is a
discussion.

\section{The clean circular corral}
\label{The clean circular corral} Starting from free electrons in 2D, we
model the boundary of an empty circular corral (without magnetic impurities
inside) by a continuous potential $W\delta (r-r_{0})$, where $r$ is the
distance to the center of the circle and $r_{0}$ its radius. The
approximation of a continuous boundary (instead of discrete adatoms forming
the corral) simplifies considerably the mathematics and is justified by the
fact that the Fermi wave length for surface electrons $2\pi /k_{F}\simeq 30$
\AA\ is larger than the average distance between adatoms $\sim 10$ \AA . The
angular momentum projection perpendicular to the surface $l_{z}=m$ becomes a
good quantum number, due to the rotational symmetry around the center of the
corral. For each $m$ and energy $E=(\hbar k)^{2}/(2m_{e}^{*})$, where $%
m_{e}^{*}$ is the effective mass, the eigenstates can be written in the form 
$\varphi _{km}(r,\theta )=\psi _{km}(kr)\ e^{im\theta }$ in polar
coordinates, and the 2D radial Schr\"{o}dinger equation becomes 
\begin{equation}
-\frac{\hbar ^{2}}{2m_{e}^{*}}\frac{1}{r}\frac{\partial \ }{\partial r}%
\left( r\frac{\partial \psi _{km}}{\partial r}\right) +[\frac{\ (\hbar m)^{2}%
}{2m_{e}^{*}r^{2}}+W\delta (r-r_{0})-E]\psi _{km}=0.  \label{sch}
\end{equation}
Its solution can be written in the form $\psi _{km}=\psi _{km}^{<}\theta
(r_{0}-r)+\psi _{km}^{>}\theta (r-r_{0})$, where $\theta (r)$ is the step
function and (except for a normalization constant): 
\begin{eqnarray}
\psi _{k,m}^{<} &=&J_{m}(kr)  \nonumber \\
\psi _{k,m}^{>} &=&A_{m}(k)J_{m}(kr)+B_{m}(k)Y_{m}(kr),  \label{psi}
\end{eqnarray}
where $J_{m}(Y_{m})$ is the Bessel function of the first (second) kind. From
the continuity of $\psi _{km}$ and the discontinuity in the first derivative
according to Eq. (\ref{sch}), and using known expressions for the
derivatives of the Bessel functions \cite{abra}, one obtains: 
\begin{eqnarray}
A_{m}(k) &=&1+\frac{2m_{e}^{*}W/(\hbar ^{2}k)}{\frac{Y_{m+1}(kr_{0})}{%
Y_{m}(kr_{0})}-\frac{J_{m+1}(kr_{0})}{J_{m}(kr_{0})}},  \nonumber \\
B_{m}(k) &=&\left( 1-A_{m}(k)\right) \frac{J_{m}(kr_{0})}{Y_{m}(kr_{0})}.
\label{ab}
\end{eqnarray}

We normalize the wave functions $\varphi _{km}(r,\theta )$ inside a hard
wall box of large radius $R$ and take the limit $R\rightarrow +\infty $. A
quantity of interest in the many-body problem (see Eq. (\ref{gf})) is the
Green's function for surface conduction electrons in the absence of the
impurity: 
\begin{equation}
G_{s}^{0}(z;r,\theta ,r^{\prime },\theta ^{\prime })=\sum_{km}\frac{\varphi
_{km}(r,\theta )\overline{\varphi }_{km}(r^{\prime },\theta ^{\prime })}{%
z-\epsilon _{km}},  \label{goc}
\end{equation}
where the bar over $\varphi _{km}$ denotes complex conjugation and the sum
over $k$ extends over all radial wave vectors for which $\varphi
_{km}(R,\theta )=0$. Using Eqs. (\ref{psi}) and (\ref{ab}), and taking the
limit $R\rightarrow +\infty $, we obtain after some algebra for $r,r^{\prime
}<r_{0}$ 
\begin{equation}
G_{s}^{0}(z;r,\theta ,r^{\prime },\theta ^{\prime })=\sum_{m}\int \frac{dk\
k\ J_{m}(kr)J_{m}(kr^{\prime })\ e^{im(\theta -\theta ^{\prime })}}{2\pi
\left( A_{m}^{2}(k)+B_{m}^{2}(k)\right) \left( z-\frac{(\hbar k)^{2}}{%
2m_{e}^{*}}\right) }.  \label{goc2}
\end{equation}
When $W\rightarrow +\infty $, it is easy to see that $A_{m},B_{m}\rightarrow
\infty $, except for those values of $k$ for which $J_{l}(kr_{0})=0$, and
one recovers the known result for the hard wall problem

\begin{equation}
G_{s}^{0}=\sum_{n,m}\frac{J_{m}(\gamma _{n}^{m}r)J_{m}(\gamma
_{n}^{m}r^{\prime })\ e^{im(\theta -\theta ^{\prime })}}{\left( \sqrt{\pi }%
r_{0}J_{m+1}(\gamma _{n}^{m}r_{0})\right) ^{2}\left( z-\frac{(\hbar \gamma
_{n}^{m})^{2}}{2m_{e}^{*}}\right) },  \label{hw}
\end{equation}
where $\gamma _{n}^{m}r_{0}$ is the $n^{th}$ zero of $J_{m}(\alpha )$. For
finite $W$ one expects that the poles of Eq. (\ref{hw}) become resonances
described by appropriate complex poles of the scattering matrix \cite
{cal,tay}. Actually, the integral of Eq. (\ref{goc2}) can be thought as an
integral over energy $E(k)=(\hbar k)^{2}/(2m_{e}^{*})$ and for any $z$ in
the upper half of the complex plane, the integrand is analytical in $E$,
except at the zeros of $A_{m}^{2}+B_{m}^{2}$. The physical zeros of $%
A_{m}^{2}+B_{m}^{2}$ lie in the lower half plane of $E$. Then in principle
one can evaluate the integral by the method of residues. A technical problem
is that the Bessel functions diverge for infinite complex $k$ and one has to
assume a high energy cut off. This however has no physical consequences.
Thus, $G_{s}^{0}$ can be written in the form: 
\begin{equation}
G_{s}^{0}=\sum_{n,m}\frac{C_{n}^{m}\
J_{m}(k_{n}^{m}r)J_{m}(k_{n}^{m}r^{\prime })\ e^{im(\theta -\theta ^{\prime
})}}{z-\epsilon _{n}^{m}+i\delta _{n}^{m}},  \label{gop}
\end{equation}
where $k_{n}^{m}$ are the complex zeros of $A_{m}^{2}(k)+B_{m}^{2}(k)$, with 
$(\hbar k_{n}^{m})^{2}/(2m_{e}^{*})=\epsilon _{n}^{m}-i\delta _{n}^{m}$, $%
\delta _{n}^{m}>0$, and from the residues of the integrand in Eq. (\ref{goc2}%
) one obtains: 
\begin{equation}
C_{n}^{m}=-\frac{i\ k_{n}^{m}}{\frac{\partial (A_{m}^{2}(k)+B_{m}^{2}(k))}{%
\partial k}|_{k=k_{n}^{m}}}.  \label{res}
\end{equation}
In practice, in Eq.(\ref{gop}) we include only the terms for which $\epsilon
_{n}^{m}<E_{c}$, where the cut off energy $E_{c}$ is typically taken as
three times the Fermi energy. Thus, Eq. (\ref{gop}) is an asymptotic
low-energy expansion. It reduces to Eq. (\ref{hw}) for a hard-wall corral.

A quantity of interest (because it is proportional to $dI/dV$ \cite
{knorr,schi}) is the local density of conduction states $\rho
_{s}^{0}(r,\theta ,\epsilon )$ which for $r<r_{0}$ is given by 
\begin{equation}
\rho _{s}^{0}(r,\theta ,\epsilon )=\lim_{\eta \rightarrow 0^{+}}\ -\frac{1}{%
\pi }%
\mathop{\rm Im}%
G_{s}^{0}(\epsilon +i\eta ;r,\theta ,r,\theta ).  \label{roo}
\end{equation}

\begin{figure}[tbp]
\centering
\includegraphics[scale=0.5,clip,keepaspectratio]{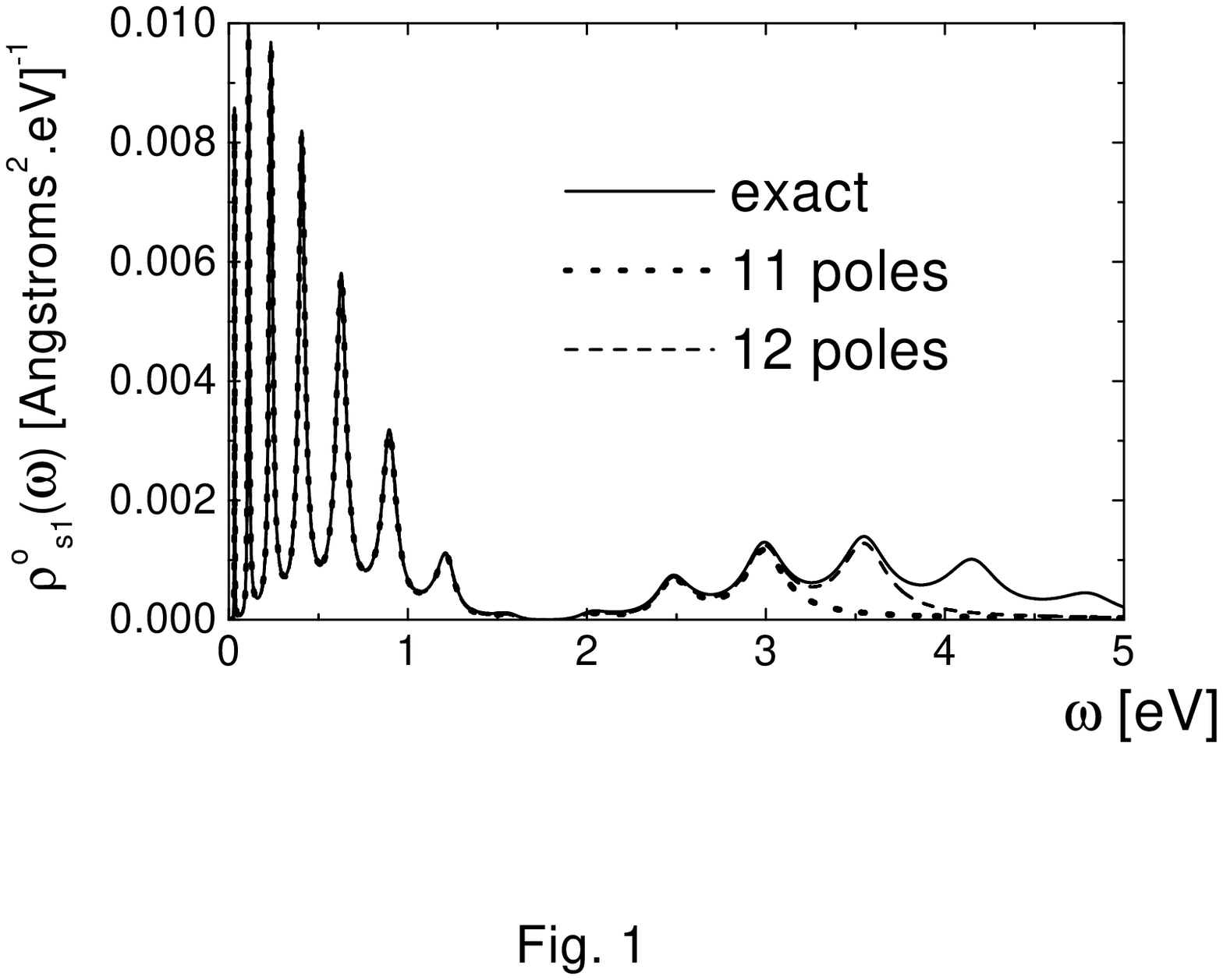}
\caption{Partial contribution of conduction states with $m=1$ to the local
density of states, and approximation using a summation of a finite number of
contributions from poles in Eqs. (\ref{gop}), (\ref{res}), as a function of
energy for $W=15 \hbar^{2}/(2m_e^{*}r_0)$ and $r=0.14\ r_0$. }
\label{fig1}
\end{figure}

In Fig. \ref{fig1} we compare the contribution to the exact $\rho _{s}^{0}$
of the states with angular momentum projection $m=1$, with the corresponding
expression using the partial summation in Eq. (\ref{gop}). We have taken an
effective mass $m_{e}^{*}=0.38\ m_{e}$ where $m_{e}$ is the electron mass 
\cite{cro,euc}. As in recent experiments \cite{man2}, we have taken $%
r_{0}=63.5$ \AA , so that the fourth state (ordered in increasing energy)
with $m=1$ falls at $\epsilon _{F}$ \cite{note}. Finally we took $W=15\hbar
^{2}/(2m_{e}^{*}r_{0})\simeq 2.37\ $eV\AA\ in order that the width of this
state would be $\delta _{4}^{1}=0.0213$ eV, similar to that reported in some
experiments \cite{cro}. The point of observation $r=0.14r_{0}$ was taken at
the maximum of $J_{1}(\gamma _{4}^{1}r)$. One can see that the result in Fig 
\ref{fig1} is consistent with the fact expected from previous studies with
potentials that vanish except in a finite region \cite{cal}: the discrete
sum Eq. (\ref{gop}) reproduces the low-energy part of $G_{s}^{0}$.

\begin{figure}[tbp]
\centering
\includegraphics[scale=0.5,clip,keepaspectratio]{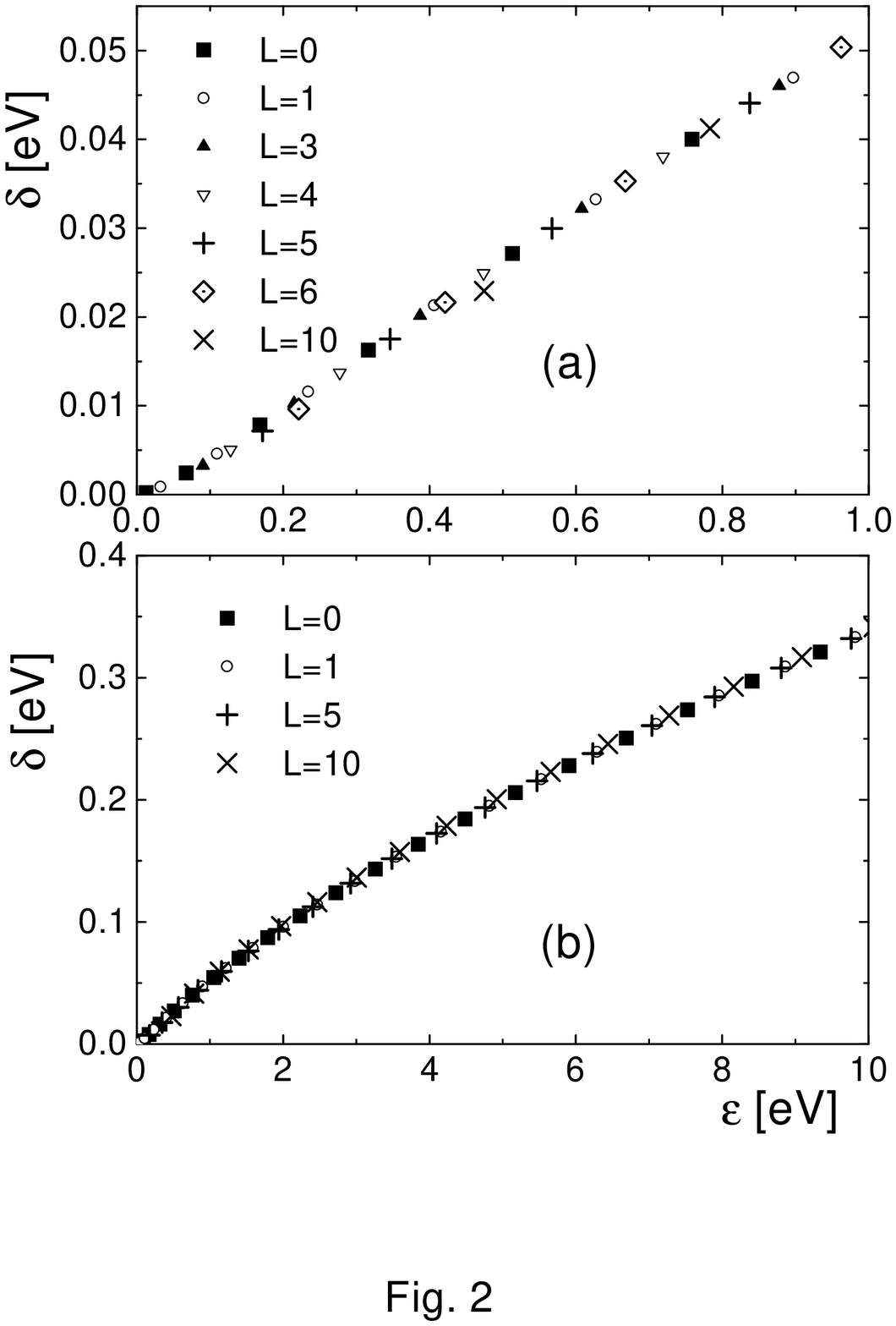}
\caption{Width of the energy levels as a function of energy, for (a) $0 <
\epsilon_{n}^{m} < 1$ eV and (b) $0 < \epsilon_{n}^{m} < 10$ eV. }
\label{fig2}
\end{figure}

In Fig. \ref{fig2} we represent $\delta _{n}^{m}$ as a function of $\epsilon
_{n}^{m}$. As a first approximation, $\delta _{n}^{m}/\epsilon _{n}^{m}$ is
a constant, independent of $m$ and $n$. In any case, $\delta _{n}^{m}$ seems
to depend on $n$ and $m$ only through $\epsilon _{n}^{m}$. This result is
surprising. For a given confining potential $W$, one would naively expect
that $\delta _{n}^{m}$ increases with the radial part of the kinetic energy,
being rather independent of the angular part. However, as $m$ increases, the
weight of the wave functions near the border of the corral ($r\sim r_{0}$)
also increases and compensates the decay in radial velocity.

To study the effects of changes in the magnitude of the confining potential,
we started form a situation with $\epsilon _{4}^{1}=\epsilon _{F}$ for rigid
walls ($W\rightarrow +\infty $). As $W$ decreases, $\epsilon _{4}^{1}$
decreases slightly, its width $\delta _{4}^{1}$ increases and the position $%
r_{M}$ of the maximum of $\rho _{s}^{0}(r,\theta ,\epsilon )$ for $\epsilon
=\epsilon _{4}^{1}$ deviates from its value $0.138r_{0}$ for $W\rightarrow
+\infty $ towards larger values. These changes are illustrated in Table \ref
{table1}. The change in $\epsilon _{4}^{1}$ can be interpreted, in a first
approximation, as a small increase of the effective mass. In fact, using
known properties of the Bessel functions \cite{abra}, it can be shown that
for $W \rightarrow\infty$, the effective mass increases by a factor $1 +
\hbar^2 /(m^*_e r_0 W)$. Fitting the lowest energies of a circular corral
with a model with hard walls, gives $m_{e}^{*}=0.38\ m_{e}$ \cite{cro,note}.
For moderate $W$, it seems that $\delta^1_4 \sim 1/W$, however for large $W$%
, $\delta^m_n \sim 1/W^2$ . In contrast to $\delta _{4}^{1}$, $r_{m}$ is
only weakly dependent on $W$ (see Table \ref{table1}).

\begin{table}[tbp]
\begin{tabular}{cccc}
\\ \hline \hline
$W\ \left[\frac{\hbar{^2}}{2m^{*}r_0}\right]$ & $\delta_4^1\ [meV]$ & $%
\epsilon_4^1\ [meV ]$ & Maximum[\% $r_o$] \\ \hline
0.1 & 171.4 & 364.8 & 14.81 \\ 
1 & 100.2 & 381.3 & 14.74 \\ 
5 & 51.0 & 393.1 & 14.60 \\ 
7 & 41.2 & 396.1 & 14.55 \\ 
10 & 31.3 & 400.7 & 14.48 \\ 
12 & 26.6 & 403.1 & 14.44 \\ 
15 & 21.3 & 406.4 & 14.39 \\ 
18 & 17.4 & 409.3 & 14.34 \\ 
$\infty$ & 0 & 441.1 & 13.80
\\ \hline \hline
\end{tabular}
\caption{Variation of the energy $\epsilon_4^1$, its width $\delta_4^1$ and
the position of the maximum of $\rho(r,\theta,\epsilon_4^1)$ as a function
of the confining potential $W$.}
\label{table1}
\end{table}

What happens with $\epsilon _{n}^{m}$ and $\delta _{n}^{m}$ if the size of
the corral is changed? We expect that $W$ remains constant, independent of $%
r_{0}$. If $r_{0}$ is increased by a factor $f$, changing variables in the
Schr\"{o}dinger equation (\ref{sch}), to $\rho =r/f$, one realizes that
(except for normalization) the following equality holds for the wave
functions of given projection, position, energy and parameters of the
potential: 
\begin{equation}
\varphi _{m}(r,\theta ,\epsilon ,fr_{0},W)\equiv \varphi _{m}(\frac{r}{f}%
,\theta ,f^{2}\epsilon ,r_{0},fW).  \label{psie}
\end{equation}
Then, increasing $r_{0}$ by a factor $f$ one expects, as a first
approximation, a radial dilation by a factor $f$ of the corresponding wave
function, a compression of the energy levels $\epsilon$ by a factor of $f^2$%
, and for large $W$, that the widths $\delta_{i}$ remain approximately
constant.

\section{The many-body problem}
In this section, we explain the model and approximations used to describe
the electronic structure of a system composed of a quantum corral and a
magnetic impurity inside it. Basically we consider surface states, described
inside the corral by a Green function $G_{s}^{0}$ like that discussed in the
previous section, hybridized with one localized $d$ orbital with an
important on-site repulsion $U$ \cite{rap,por,hal,ali2}. We also include an
hybridization of the $d$ orbital with bulk states. We are neglecting the
degeneracy of the $d$ level, which we believe is not important for the
essential physics, and other $s,p$ orbitals brought by the impurity which
might affect the magnitude of $dI/dV$ near the site impurity.

The Hamiltonian can be written as: 

\begin{eqnarray}
H&=&\sum_{j\sigma }\varepsilon _{j}s_{j\sigma }^{\dagger }s_{j\sigma
}+\sum_{j\sigma }\varepsilon _{j}^{b}b_{j\sigma }^{\dagger }b_{j\sigma
}+E_{d}\sum_{i\sigma }d_{i\sigma }^{\dagger }d_{i\sigma }+ \nonumber
\\
&&+U\sum d_{\uparrow
}^{\dagger }d_{\uparrow }d_{\downarrow }^{\dagger }d_{\downarrow }  \nonumber
+\sum_{j\sigma }V_{s}\lambda [\varphi _{j}(R_{i})d_{\sigma }^{\dagger
}s_{j\sigma }+\text{H.c.}]+\nonumber\\
&&+\sum_{j\sigma }V_{b}^{j}[d_{\sigma }^{\dagger
}b_{j\sigma }+\text{H.c.}].  \label{ham}
\end{eqnarray}
where $s_{j\sigma }^{\dagger }$ ($b_{j\sigma }^{\dagger }$) are creation
operators for an electron in the $j^{th}$ surface (bulk) conduction
eigenstate in the absence of the impurity but including the corral. The
impurity is placed at the two-dimensional position $R_{i}$ on the surface,
and we assume that the hybridization of the impurity $d$ orbital with the
surface state $j$ is proportional to its normalized wave function at that
point $\varphi _{j}(R_{i})$. We write the proportionality constant as $%
V_{s}\lambda $, where $V_{s}$ is an energy (representing the hybridization
in a tight binding model \cite{wei,ali2}) and $\lambda =2.38$ \AA\ is the
square root of the surface per Cu atom of a Cu(111) surface.

At sufficiently low temperature, the differential conductance $dI/dV$ when
the STM tip is at position $R_{t}$ is proportional to the density of the
mixed state \cite{schi}, 
\begin{equation}
f_{\sigma }(R_{t})=\lambda \sum_{j}\varphi _{j}(R_{t})s_{j\sigma
}+qd_{\sigma },  \label{q}
\end{equation}
\begin{equation}
\frac{dI(V)}{dV}\sim \rho _{f}(\epsilon _{F}+eV)=-\frac{1}{\pi }%
\mathop{\rm Im}%
G_{f}^{\sigma }(\epsilon _{F}+eV),  \label{didv}
\end{equation}
where $G_{f}^{\sigma }(\omega )=\langle \langle f_{\sigma };f_{\sigma
}^{\dagger }\rangle \rangle _{\omega }$ is the Green function of $f_{\sigma
}(R_{t})$, and $q(R_{t})$ is the ratio between the tunneling matrix elements
tip-impurity and tip-surface, and is significant only for very small $%
R_{t}-R_{i}$, due to the rapid decay of the wave functions for $d$ states.

Using the equations of motion 
\begin{eqnarray}
(\omega -\epsilon _{j})\langle \langle s_{j\sigma };s_{j\prime \sigma
}^{\dagger }\rangle \rangle _{\omega } &=&\delta _{jj\prime }+V_{s}\lambda 
\overline{\varphi }_{j}(R_{i})\langle \langle d_{\sigma };s_{j\prime ,\sigma
}^{\dagger }\rangle \rangle _{\omega },  \nonumber \\
(\omega -\epsilon _{j})\langle \langle s_{j\sigma };d_{\sigma }^{\dagger
}\rangle \rangle _{\omega } &=&V_{s}\lambda \overline{\varphi }%
_{j}(R_{i})\langle \langle d_{\sigma };d_{\sigma }^{\dagger }\rangle \rangle
_{\omega },  \nonumber \\
(\omega -\epsilon _{j\prime })\langle \langle d_{\sigma };s_{j\prime \sigma
}^{\dagger }\rangle \rangle _{\omega } &=&V_{s}\lambda \varphi _{j\prime
}(R_{i})\langle \langle d_{\sigma };d_{\sigma }^{\dagger }\rangle \rangle
_{\omega },  \label{m}
\end{eqnarray}
$G_{f}^{\sigma }$ is expressed in terms of the Green function for the $d$
electrons $G_{d}^{\sigma }(\omega )=\langle \langle d_{\sigma };d_{\sigma
}^{\dagger }\rangle \rangle _{\omega }$, and the unperturbed conduction
electron Green function for surface states $G_{s}^{0}$ . We drop the
superscript $\sigma $ in the following because of the spin independence of
the problem. The difference in $G_{f}$ between the results with and without
impurity becomes: 
\begin{eqnarray}
\Delta G_{f}=\left( V_{s}\lambda ^{2}G_{s}^{0}(R_{t},R_{i},\omega
)+q\right) \times  \nonumber \\
\times \left( V_{s}\lambda ^{2}G_{s}^{0}(R_{i},R_{t},\omega )+q\right) G_{d},.
\label{gf}
\end{eqnarray}
where 
\begin{equation}
G_{s}^{0}(R_{1},R_{2},\omega )=\sum_{j}\frac{\varphi _{j}(R_{1})\overline{%
\varphi }_{j}(R_{2})}{\omega -\epsilon _{j}}.  \label{gs}
\end{equation}
We assume that $G_{s}^{0}$ is known from the one-body problem. For the
circular corral, it has been discussed in the previous section.

The remaining task is to calculate $G_{d}$ using a many-body approach. The
most accurate calculation would be to use the Wilson renormalization group 
\cite{wrg}. However, the particular structure of the one-body problem, and
the lack of continuous symmetries in the general case renders its
application very difficult. We have used perturbation theory up to second
order in $U$ \cite{yos,hor}. Near the symmetric case $E_{d}+U/2=\epsilon
_{F} $, the theory is quantitatively correct up to $U\sim 8\Delta $, where $%
\Delta $ is the resonant level width \cite{sil}. Out of the symmetric case,
interpolative self-consistent schemes \cite{lev,kaj} still work for
moderately large values of $U$ \cite{dots,pc}. In particular, the persistent
current in small rings with quantum dots practically coincides with exact
results \cite{pc}. Here we assume a situation in which the effective
one-particle $d$ level $E_{d}^{eff}=E_{d}+U\langle d_{\sigma }^{\dagger
}d_{\sigma }\rangle $ is very near the Fermi level. This is justified by
first-principle calculations \cite{llois} and allows us to avoid
selfconsistency.

Then: 
\begin{equation}
\lbrack G_{d}(\omega )]^{-1}=[G_{d}^{0}(\omega )]^{-1}-\Sigma (\omega ),
\label{gd}
\end{equation}
where $G_{d}^{0}$ is $G_{d}$ for $U=0$ and $E_{d}$ replaced by $E_{d}^{eff}$%
, and $\Sigma (\omega )$ is the self-energy up to second order in $U$
evaluated from a Feynmann diagram involving the analytical extension of the
time ordered $G_{d}^{0}$ to Matsubara frequencies \cite{yos,hor}: 
\begin{eqnarray}
\Sigma (i\omega _{l},T) &=&U^{2}T\sum_{m}G_{d}^{0}(i\omega _{l}-i\nu
_{m})\chi (i\nu _{m}),  \label{sigma} \\
\chi (i\nu _{m}) &=&-T\sum_{n}G_{d}^{0}(i\omega _{n})G_{d}^{0}(i\omega
_{n}+i\nu _{m}),  \label{xi}
\end{eqnarray}
where $\omega _{n}=(2n+1)\pi T$ and $\nu _{m}=2n\pi T$.

From the equation of motion of $G_{d}$ for $U=0$ and using Eqs. (\ref{m}) we
obtain 
\begin{equation}
G_{d}^{0}(\omega )=\frac{1}{\omega -E_{d}^{eff}+i\delta _{b}-(V\lambda
)^{2}G_{s}^{0}(R_{i},R_{i},\omega )},  \label{god}
\end{equation}
where we have replaced 
\begin{equation}
\text{Im}\sum_{i}\frac{|V_{b}^{i}|^{2}}{\omega +i\eta -\epsilon _{i}}%
=-\delta _{b}  \label{dimp}
\end{equation}
assuming that the $V_{b}^{i}$ and the bulk density of states are featureless
near $\epsilon _{F}$ and can be replaced by a constant. The real part of the
sum can be absorbed in $E_{d}^{eff}$.

From the discussion of section \ref{The clean circular corral}, and previous
works \cite{cal}, we know that $G_{s}^{0}(R_{i},R_{i},\omega )$ can be
expanded as a sum of contributions from discrete poles. In particular, for
the circular corral, $G_{s}^{0}(\omega ;r,\theta ,r^{\prime },\theta
^{\prime })$ is given by Eqs. (\ref{gop}) and (\ref{res}). For other shapes
of the corral, the results of the previous section suggest that within
errors of a few percent: 
\begin{equation}
G_{s}^{0}(\omega ,R_{1},R_{2})\simeq \sum_{j}\frac{\varphi _{j}^{c}(R_{1})%
\overline{\varphi }_{j}^{c}(R_{2})}{\omega -\epsilon _{j}+i\delta
_{F}\epsilon _{j}/\epsilon _{F}},  \label{goap}
\end{equation}
where now $\varphi _{j}^{c}(R)$ are the discrete eigenstates of the hard
wall corral and $\epsilon _{j}$ are their energies, calculated with a
slightly renormalized mass. $\delta _{F}$ is the width of the resonance at
the Fermi level. In practice we cut off the sum in Eq. (\ref{goap}) at an
energy $\sim 3\epsilon _{F}$, retaining $N$ poles. Then, $G_{d}^{0}(\omega )$
can be expressed as a sum of $N+1$ poles and residues at these poles \cite
{notepol} Denoting the poles of $G_{d}^{0}(\omega )$ by $r_{l}-i\delta _{l}$
and their residues by $a_{l}$, we can write: 
\begin{equation}
G_{d}^{0}(\omega )=\sum_{l=1}^{N+1}\frac{a_{l}}{\omega -r_{l}+i\delta _{l}}.
\label{gdp}
\end{equation}
This is the retarded Green function. In order to evaluate perturbative
diagrams, like Eqs. (\ref{sigma}), (\ref{xi}), one needs the analytical
extension of the time ordered Green function to imaginary frequencies $%
\omega \rightarrow i\omega _{n}$ \cite{mahan,note2}: 
\begin{equation}
G_{d}^{0}(i\omega _{n})=\sum_{l}\frac{a_{l}+\overline{a_{l}}+(a_{l}-%
\overline{a_{l}})\text{sgn}(\omega _{n})}{2\left[ i\omega _{n}+i\delta _{l}%
\text{sgn}(\omega _{n})-r_{l}\right] },  \label{gdn}
\end{equation}
where $\overline{a_{l}}$ is the complex conjugate of $a_{l}$ and sgn$(x)$ is
the sign of $x$. Here and in what follows, the origin of energies has been
placed at the Fermi energy $(\epsilon _{F}=0)$ for simplicity.

Using standard methods \cite{mahan}, the sums over Matsubara frequencies,
Eqs. (\ref{sigma}), (\ref{xi}) can be transformed into integrals over branch
cuts. The first integral can be done analytically in terms of the digamma
function $\Psi (z)$ \cite{abra}. A lengthy but straightforward algebra after
analytical continuation back to real $\omega $ leads to: 
\begin{eqnarray}
\Sigma (\omega )&=&\frac{U^{2}}{\pi }\int_{-\infty }^{\infty }dy\ [\frac{1}{
e^{y/T}-1}G_{d}^{0}(\omega -y)\mathop{\rm Im}
\chi (y)+\nonumber \\
&&+\frac{1}{e^{y/T}+1}\text{Im}G_{d}^{0}(-y)\ \chi (\omega +y)],
\label{sigf}
\end{eqnarray}

where 
\begin{eqnarray}
\chi (\omega ) &=&-\frac{i}{2\pi }\ \sum_{lm}\left( A_{lm}(\omega
)-B_{lm}(\omega )\right) \Psi _{l}(0)+\nonumber \\
&&+\left( \overline{B_{lm}}(-\omega )-
\overline{A_{lm}}(-\omega )\right) \overline{\Psi _{l}}(0)+  \nonumber \\
&&-\left( B_{lm}(\omega )+A_{lm}(-\omega )\right) \Psi _{l}(\omega
)-\nonumber \\
&&-\left( 
\overline{B_{lm}}(-\omega )+\overline{A_{lm}}(\omega )\right) \overline{\Psi
_{l}}(-\omega ),  \label{xif}
\end{eqnarray}

with

\begin{eqnarray}
A_{lm}(\omega ) &=&\frac{a_{l}\ a_{m}}{\omega +r_{l}-r_{m}-i\delta
_{l}+i\delta _{m}},  \nonumber \\
B_{lm}(\omega ) &=&\frac{a_{l}\ \overline{a_{m}}}{\omega
-r_{l}+r_{m}+i\delta _{l}+i\delta _{m}},  \nonumber \\
\Psi _{l}(\omega ) &=&\Psi \left( \frac{1}{2}+\frac{\delta
_{l}+i(r_{l}-\omega )}{2\pi T}\right) .  \label{aux}
\end{eqnarray}
Summarizing, our approximation for the Green function of the $d$ electrons $%
G_{d}$ consist in the following steps: we first decompose the unperturbed
Green function $G_{d}^{0}$ given by Eq. (\ref{god}), and (\ref{gop}) or (\ref
{goap}) into a sum of poles, Eq. (\ref{gdp}). Then, the self-energy $\Sigma
(\omega )$ is calculated using Eqs. (\ref{sigf}), (\ref{xif}) and (\ref{aux}%
), and finally $G_{d}$ is obtained from Eq. (\ref{gd}). The change in $dI/dV$
after adding the impurity can then be evaluated using Eqs. (\ref{didv}) and (%
\ref{gf}).

\section{Results}

\subsection{The elliptical corral}
\begin{figure}[tbp]
\centering
\includegraphics[scale=0.5,clip,keepaspectratio]{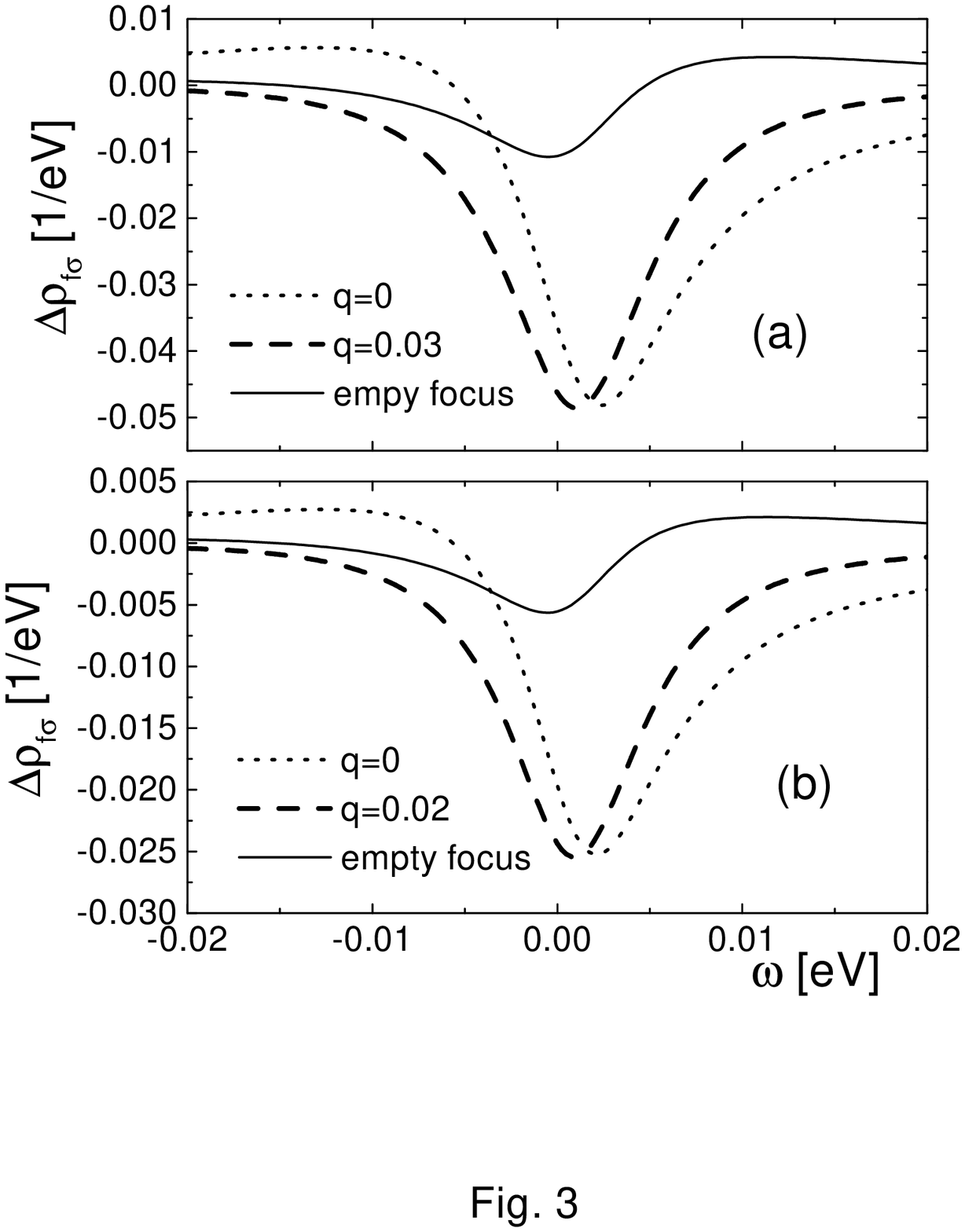}
\caption{Change in the spectral density of the $f$ state, proportional to $%
\Delta dI/dV$ (see Eqs. (\ref{didv}), (\ref{gf})) as a function of frequency
at the impurity site for two values of $q$ and at the other focus, for two
sets of parameters (a) $V_{s}=0.68$ eV, $\delta_b=0$, $E_{d}^{eff}=0.015$ eV
(b) $V_{s}=0.48$ eV, $\delta_b=0.032$ eV, $E_{d}^{eff}=0.007$ eV. In both
cases $\delta_{F}=0.04$ eV, $U=1$ eV, $T=0$.}
\label{fig3}
\end{figure}

The line shape of the change in differential conductance after adding the
impurity, $\Delta dI/dV$ has been reported for the elliptical corral with
eccentricity $e=0.5$ and semimajor axis of $71.3$ \AA\ with a Co impurity
placed at the left focus \cite{man}. Comparison with our results determines
the value of $V_{s}$ which leads to the observed line width $\sim 0.009$ eV.
We took $U=1\ $eV and assumed a situation near the symmetric case, for which
the Hartree-Fock $E_{d}^{eff}$ level is near $\epsilon _{F}$ \cite{llois}.
We shifted $E_{d}^{eff}$ so that the minimum of $\Delta dI/dV$ as a function
of energy lies near the experimental position. $E_{d}^{eff}$ might be
calculated using self-consistent interpolative approaches (\cite
{lev,kaj,dots,pc}), but we avoided them here. We have taken two values for
the level width of the conduction states at $\epsilon _{F}$, $\delta
_{F}=0.02$ eV, suggested by some experiments \cite{cro} and $\delta
_{F}=0.04 $ eV, which leads to the observed ratio $\sim 1/8$ between the
intensity in $dI/dV$ at the mirage ($-R_{i}$) and at the impurity ($R_{i}$) 
\cite{rap}. For $\delta _{F}=0.04$ eV and $\delta _{b}=0$, we obtain that $%
V_{s}=0.68$ eV and $q\sim 0.03$, leads approximately to the observed line
shape. They also explain the line shape when the impurity is placed on a
clean surface \cite{ali2}. We then decrease the value of $V_{s}$ by a factor
of $\sqrt{2}$ to $V_{s}=0.48$ eV and increase the resonant level width of
the impurity to $\delta _{b}=0.032$ eV in such a way that the width of the
Kondo peak in the impurity density of states $\rho _{d}(\omega )$ (not
shown) is the same as before. This conditions implies approximately the same
contribution from bulk and surface states to the Kondo resonance, as assumed
in previous studies \cite{aga}. As shown in Fig. \ref{fig3}, and in
agreement with previous calculations \cite{wil}, both sets of parameters
lead to very similar results except for a scale factor. This factor is
approximately $V_{s}^{2}$, as expected from Eq. (\ref{gf})

\begin{figure}[tbp]
\centering
\includegraphics[scale=0.5,clip,keepaspectratio]{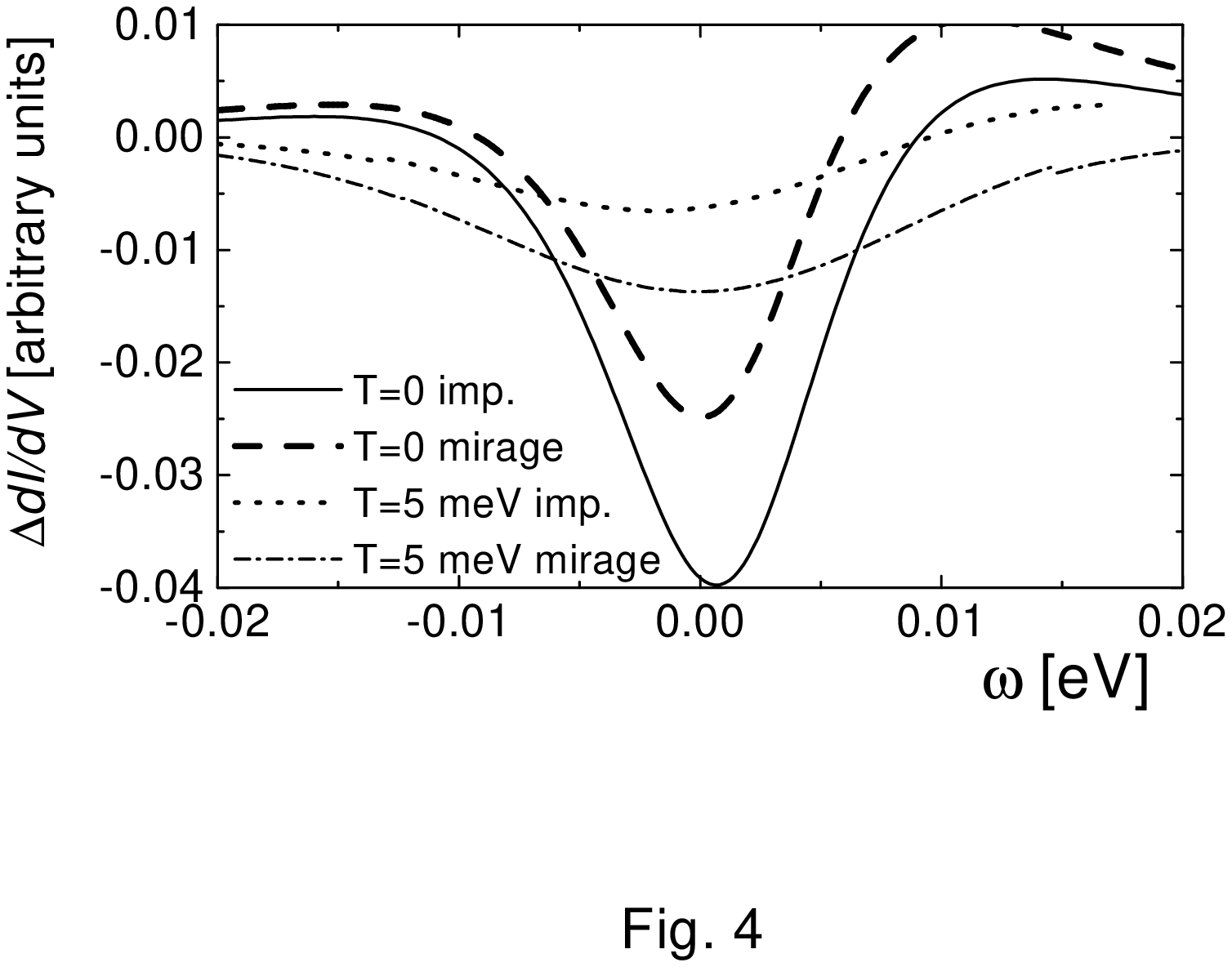}
\caption{Change in differential conductance at both foci for two
temperatures. Parameters are $V_{s}=0.45$ eV, $\delta_{F}=0.02$ eV, $%
\delta_b=0.035$ eV, $E_{d}^{eff}=0.007$ eV, $U=1$ eV, $q=0.02$.}
\label{fig4}
\end{figure}

The decrease in the intensity of the resonance at the mirage point with
respect to the impurity position can be understood from Eqs. (\ref{gf}) and (%
\ref{goap}) and is a consequence of the destructive interference of the $%
42^{th}$ state of the ellipse (which lies at the Fermi level and is even
under reflection through the minor axis $\sigma $) with other states which
are odd under $\sigma $ (like states 32, 35, and 51): all terms contribute
to the same sign to the dominant imaginary part of the sum in Eq. (\ref{goap}%
) for $R_{1}=R_{2}$, while there is a partial cancellation for $R_{1}=-R_{2}$%
. This interference decreases with decreasing $\delta _{F}$ and the
intensity at the mirage point increases. For $\delta _{F}=0.02$ eV, $V_{s}$
should be decreased slightly to $0.63$ eV to keep the same width of the
resonance, for $\delta _{b}=0$. Taking $V_{s}=0.45$ eV and increasing $%
\delta _{b}$ from zero to $\delta _{b}=0.035$ eV to keep the width of $\rho
_{d}(\omega )$ (as above), the result for $\Delta \int d\omega \rho
_{f}(\omega )(-\partial f(\omega )/\partial \omega )\sim \Delta dI/dV$ is
shown in Fig. \ref{fig4} for two different temperatures. At $T=0$, in
comparison with the previous results for $\delta _{F}=0.04$ eV, the
intensity at the mirage increased to nearly $60\%$ of that at the impurity
position. At the Kondo temperature, the resonances at both positions broaden
and the amplitude at the minima decrease more than $50\%$ of the zero
temperature result. Most of the broadening is due to the effect of the
derivative of the Fermi function, $-\partial f(\omega )/\partial \omega $.
In fact, $\Delta \rho _{f}(\omega )$ at the impurity looses only $\sim 1/3$
of the intensity and broadens very little as $T$ is increased from $0$ to $%
0.005\ eV$.

\begin{figure}[tbp]
\centering
\includegraphics[scale=0.5,clip,keepaspectratio]{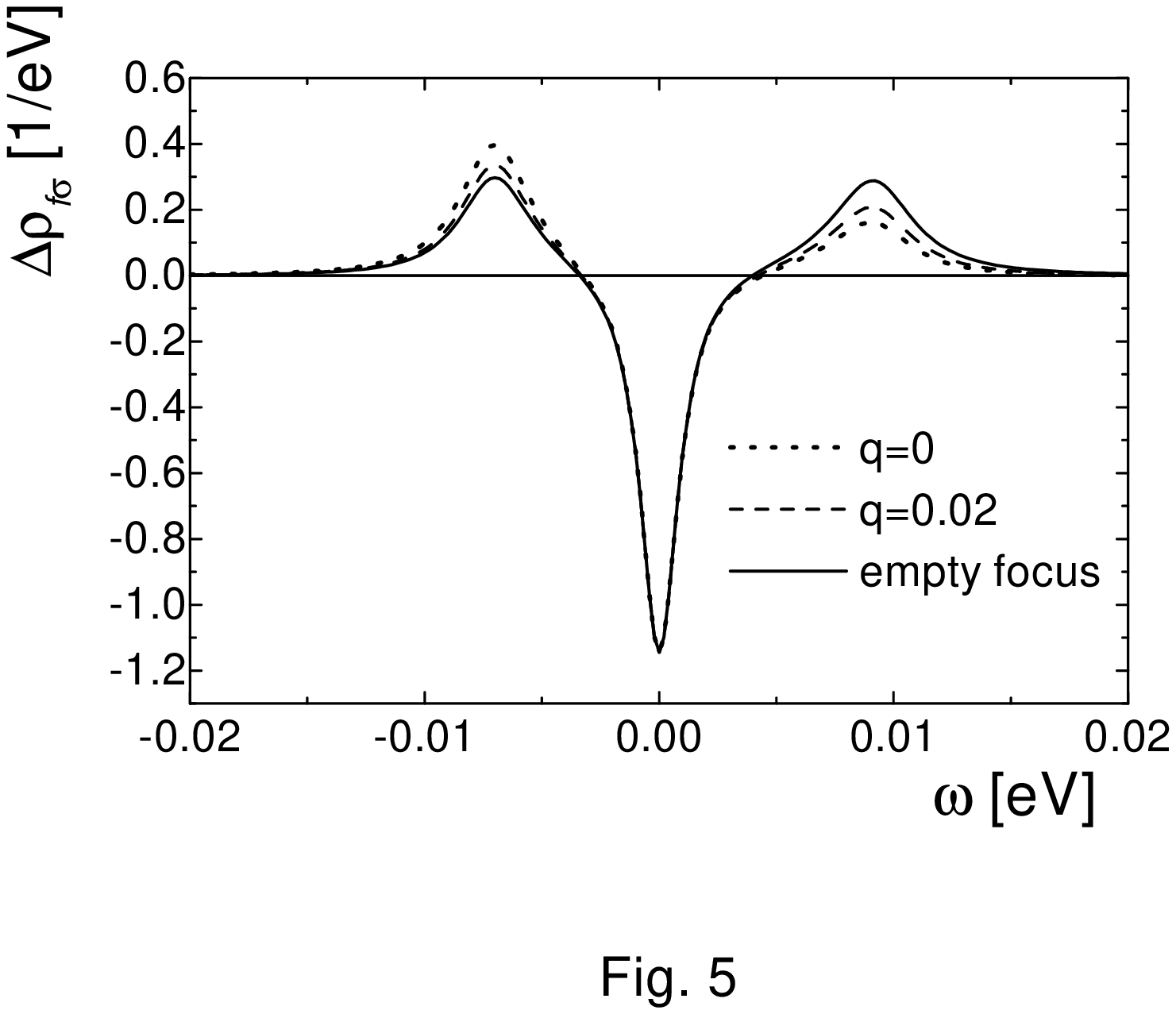}
\caption{$\Delta \rho _{f\sigma}\sim \Delta dI/dV$ as a function of
frequency for $\delta_F= 0.001$ eV and other parameters as in Fig. \ref{fig4}%
.}
\label{fig5}
\end{figure}

\begin{figure}[tbp]
\centering
\includegraphics[scale=0.5,clip,keepaspectratio]{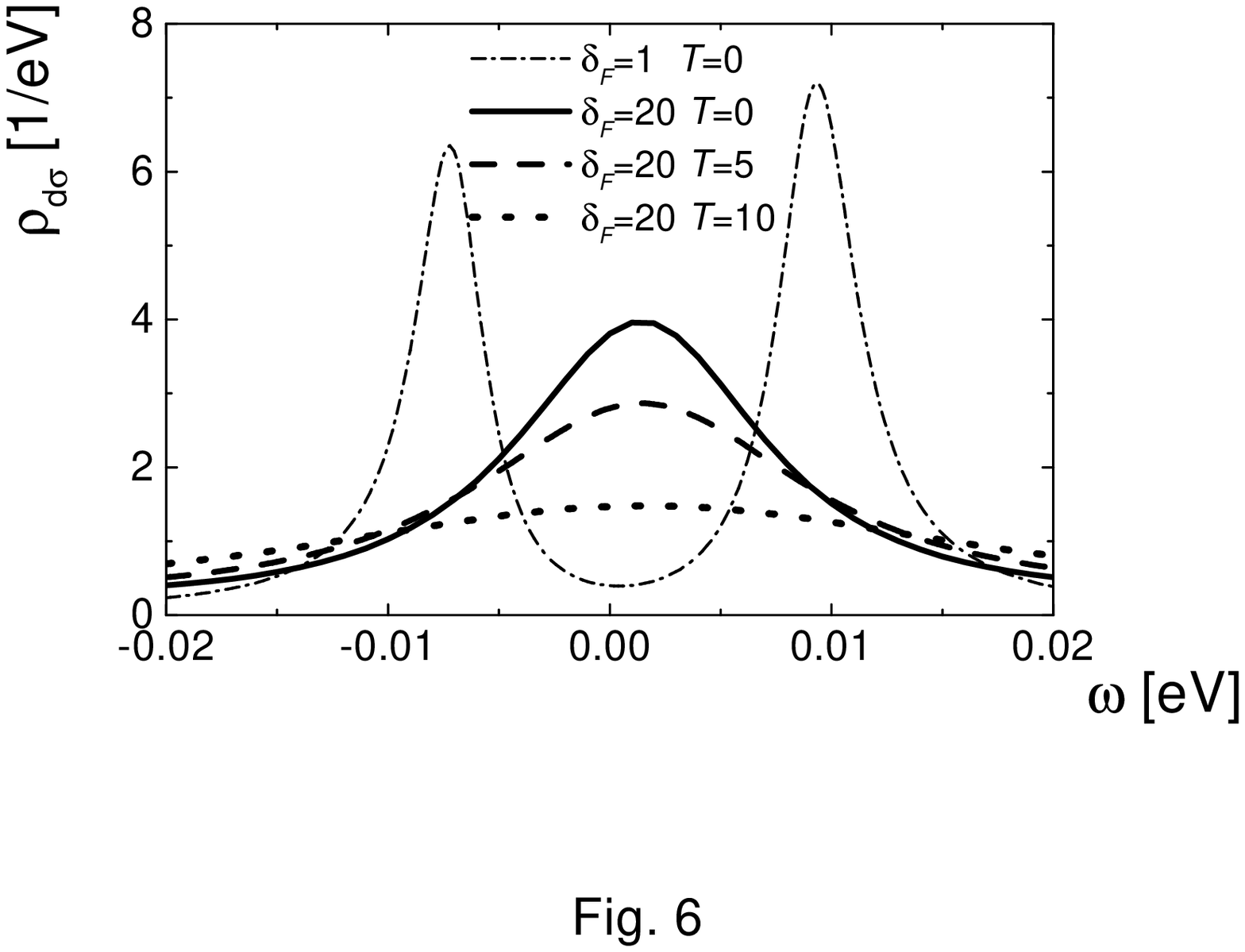}
\caption{Impurity spectral density for several values of $\delta_{F}$ and $T$
indicated in meV inside the figure. Other parameters as in Fig. \ref{fig4} }
\label{fig6}
\end{figure}

If $\delta _{F}$ is further reduced, the space dependence of $\Delta \rho
_{f}(\omega )$ is almost exactly that of the conduction eigenstate at $%
\epsilon _{F}$ and therefore the intensities at both foci are practically
the same. However, the shape of the resonance strongly changes, and two
positive contributions to $\Delta \rho _{f}(\omega )$ appear at both sides
of $\epsilon _{F}$. This has been shown before for $\delta _{b}=0$, \cite
{ali2} and is a consequence of the splitting of the Kondo peak in the
impurity density of states $\rho _{d}(\omega )$ \cite{rap,wil}. This
splitting of the Kondo peak into two peaks at both sides of the Fermi
energy, which was obtained first in perturbation theory \cite{rap,ali2} and
later by exact calculation of a reduced Hamiltonian plus embedding in the
rest of the system \cite{wil} has been recently confirmed by numerical
Wilson renormalization group calculations in a simpler system with U(1)
symmetry \cite{cor}. As shown in Fig. \ref{fig5}, this effect remains for $%
\delta _{b}\ne 0$, although the split peaks in $\rho _{d}(\omega )$ do not
reduce to two delta functions for $\delta _{F}=0$ in this case (see Fig. \ref{fig6}). 
In Fig. \ref{fig6} we also show the temperature dependence of $\rho
_{d}(\omega )$ for the parameters of Fig. \ref{fig4}.

So far, the experiments were done at very small temperatures ($T\cong 4$ K $%
\ll T_{K}\cong 50$ K), and there is no appreciable difference with the $T=0$
results. A study of the temperature dependence would confirm that the
resonance is due to a many-body effect. In general (including other
situations discussed below), the scale of the $T$ dependence is determined
by $T_{K}$, which in turn is given by half of the with of the peak in the 
{\em impurity} spectral density $\rho _{d}(\omega )$. As can be seen
comparing Figs. 4 and 6, for moderate $\delta _{F}$ (like $\delta _{F}=0.02$
eV), this width is larger than the width of the dip in the {\em conduction
electron} spectral density $\rho _{c}(\omega )$ or in $\rho _{f}(\omega )$.

The essential features of the space dependence of $\Delta dI/dV$ are very
similar to that discussed previously \cite{rap,wil} and will not be
reproduced here. Essentially it reflects the electron density of the
conduction state at $\epsilon _{F}$, somewhat blurred out of the impurity
position for $\delta _{F}\gtrsim 0.03$\ eV.

\begin{figure}[tbp]
\centering
\includegraphics[scale=0.5,clip,keepaspectratio]{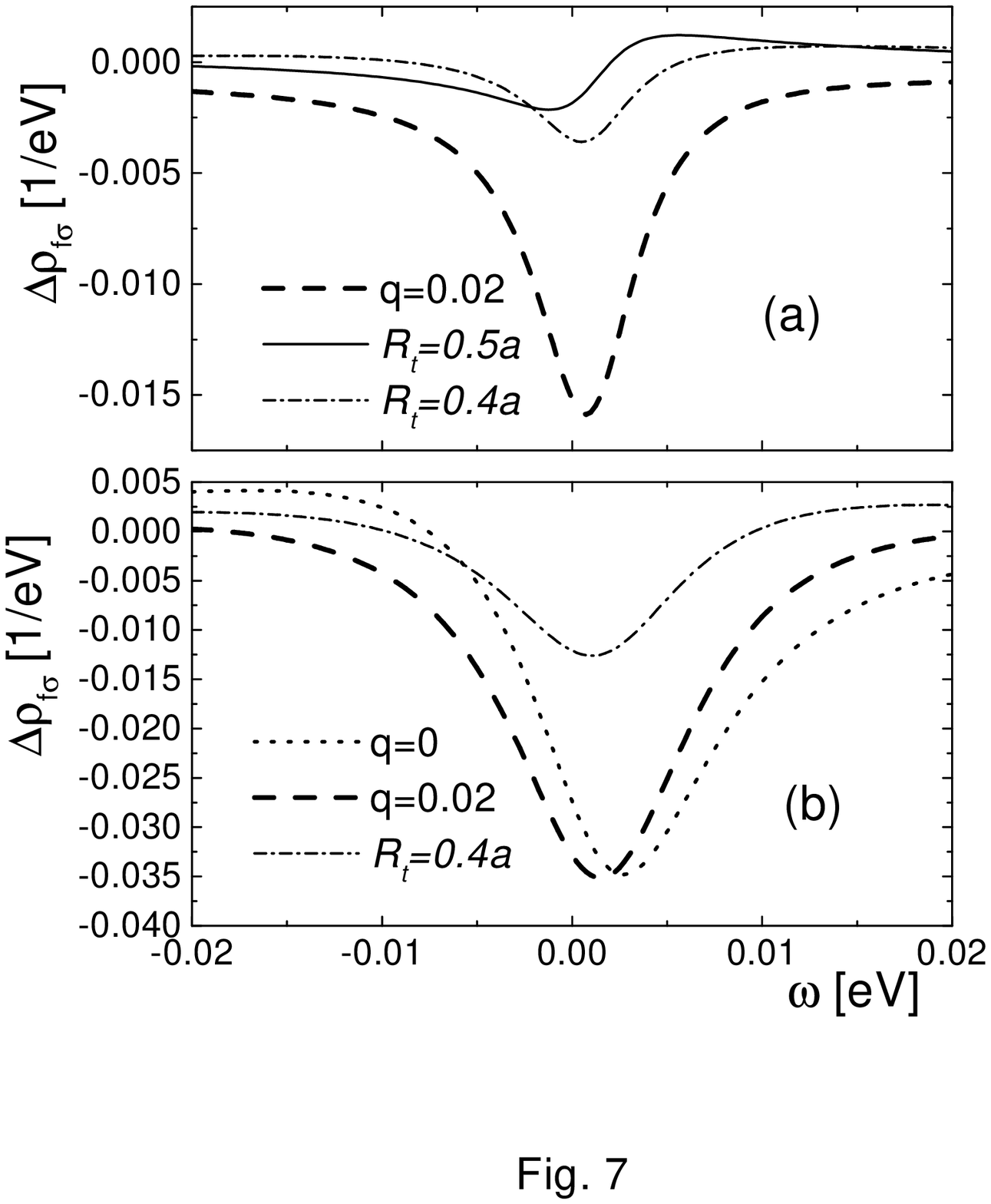}
\caption{$\Delta \rho _{f\sigma}\sim \Delta dI/dV$ as a function of
frequency at several points for an ellipse with $a=64.6$\ \AA and two
impurity positions: (a) $R_{i}=-0.5a$ and (b) $R_{i}=-0.4a$. Other
parameters as in Fig \ref{fig4}. }
\label{fig7}
\end{figure}

In the following we use the four set of parameters which lead to a
reasonable line shape at the impurity site ($\delta _{F}=0.02$\ eV or $%
\delta _{F}=0.04$\ eV, $\delta _{b}=0$ or $\delta _{b}\ne 0$) to study other
situations for which the voltage dependence of $\Delta dI/dV$ has not been
reported. We first consider an elliptical corral with the same eccentricity
but semimajor axis reduced from $a=71.3$ \AA\ to $a=64.6$ \AA . In this
case, the conduction state 35 (instead of 42) of the ellipse falls at $%
\epsilon _{F}$. The electronic density of this state $\left| \varphi
_{35}(R)\right| ^{2}$, has maxima at $R=\pm 0.4a$, displaced from the foci $%
R=\pm 0.5a$. Studying the negative interference at the mirage point, one
expects a stronger mirage at $R_{t}=0.4a$ if the impurity is placed at $%
R_{i}=-0.4a$ \cite{rap}. We studied two cases, with the impurity at one
focus or at one maximum of $\left| \varphi _{35}(R)\right| ^{2}$ for the
four set of parameters mentioned above (actually, from the discussion of the
previous section, one would expect a $10\%$ reduction of $\delta _{F}$, with
respect to the larger ellipse, but we neglect it). For $R_{i}=-0.5a$
(impurity on focus), the resonance narrows with respect to the previous
case. This is mainly due to the fact that the effective hybridization of the
state at the Fermi level is proportional to $\left| \varphi
_{35}(R_{i})\right| ^{2}$ which is approximately $60\%$ of $\left| \varphi
_{42}(R_{i})\right| ^{2}$ of the larger ellipse. The total width is near $%
0.005$ eV for $\delta _{b}=0$ , and near $0.007$\ eV for $\delta _{b}\ne 0$
(see Fig. \ref{fig7} (a)). The intensity of the depression at the other
focus $R_{t}=0.5a$ is roughly half of that at $R_{t}=0.4a$ at the extremum
of $\varphi _{35}(R_{i})$ near that focus. In particular for $\delta
_{F}=0.02$\ eV (not shown), the dip in $\Delta \rho _{f}(\omega )$ at $%
R_{t}=0.4a$ is slightly more pronounced than that at the impurity position $%
R_{t}=R_{i}=-0.5a$.

When the impurity is placed at one extremum of $\varphi _{35}(R)$, $%
R_{t}=0.4a$, the peak broadens with respect to the larger ellipse to $\sim
0.015$\ eV for $\delta _{b}=0$, or $0.012$\ eV for $\delta _{b}\ne 0$, due
to the $\sim 20\%$ larger effective hybridization with the state at $%
\epsilon _{F}$ (see Fig. \ref{fig7} (b)). As expected \cite{rap}, the
intensity at the mirage point $R_{t}=0.4a$, compared to the impurity point
is larger in the smaller ellipse.

While the {\em intensity} of the mirage is closely related with the
hybridization of the impurity with the state at $\epsilon _{F}$, the {\em %
width} of the peak in the impurity spectral density and hence $T_{K}$
depends on the effective hybridization with the impurity of all conduction
states. For example, if in the ellipse of the original experiments one
suppresses artificially the hybridization with the main odd states 32, 35,
and 51, the width in $\rho _{d}(\omega )$ is reduced to roughly half its
value for realistic $\delta _{F}$, while the magnitude of $\Delta dI/dV$ at
the mirage point {\em increases} substantially, reaching almost the same
magnitude as in the focus occupied by the impurity. The width in $\Delta
dI/dV$ is less sensitive to the parameters as that of $\rho _{d}(\omega )$.
As expected, this sensitivity is decreased for increasing $\delta _{F}$.

\subsection{The circular corral}
In this section we study the voltage and space dependence of $\Delta dI/dV$
for a circular corral, for conditions corresponding to recent experiments 
\cite{man2}: a radius $r_{0}\sim 63.5$ \AA , such that the degenerate $%
37^{th}$ and $38^{th}$ conduction eigenstates lie at the Fermi energy. In
contrast to the elliptical corral, we do not use the approximate expression (%
\ref{goap}) for the conduction electron Green's function, but the correct
expansion given by Eqs. (\ref{gop}) and (\ref{res}). The width of the
resonant states is determined by the potential scattering at the boundary $W$%
, and its value at the Fermi energy $\delta _{F}$ is given in Table \ref{table1}.

\begin{figure}[tbp]
\centering
\includegraphics[scale=0.5,clip,keepaspectratio]{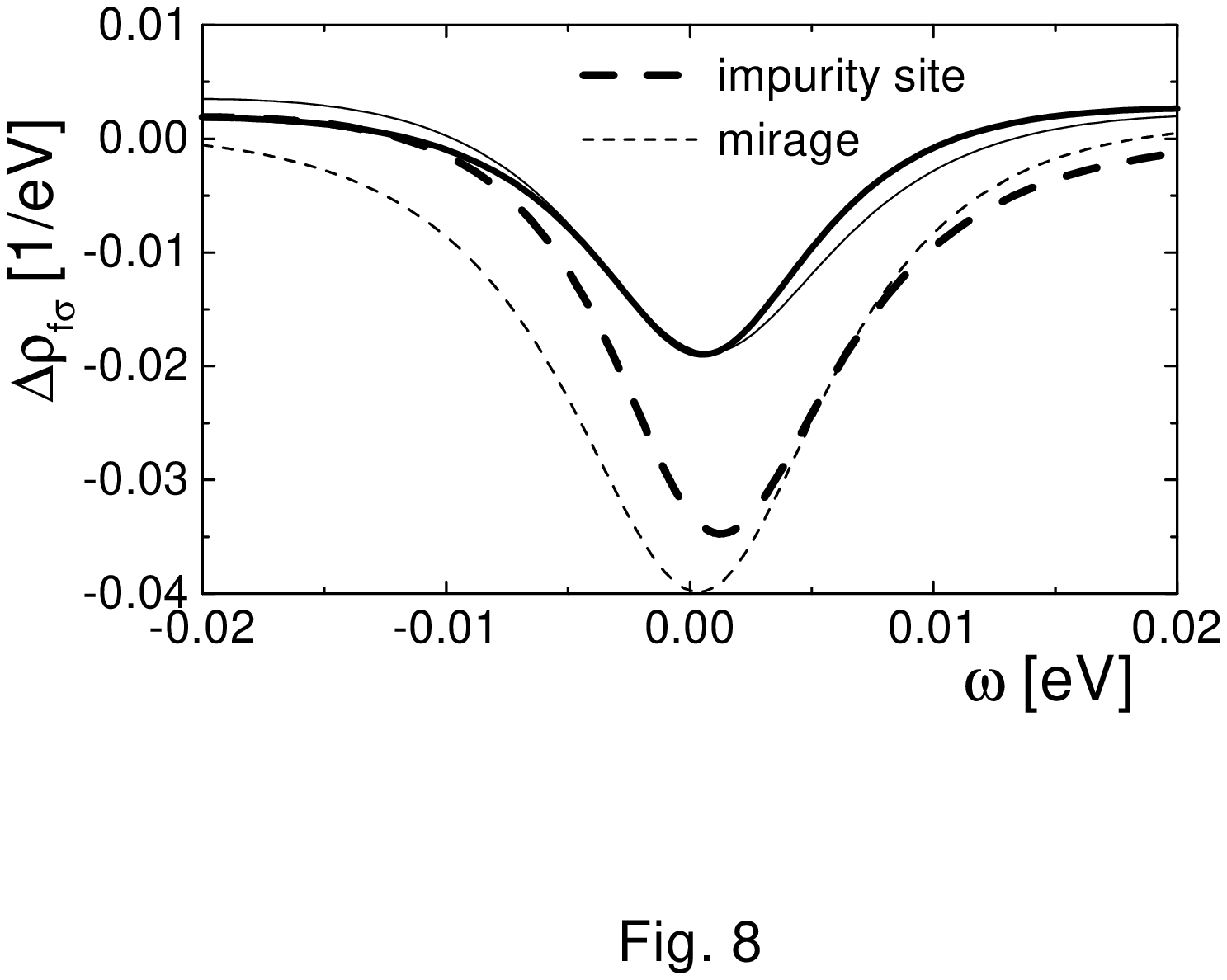}
\caption{$\Delta \rho _{f\sigma}\sim \Delta dI/dV$ as a function of
frequency for $R_{t}=R_{i}=0.146 r_0$ (dashed line) and $R_{t}=- 0.18 r_0$
(solid line) for the circular corral. The thinner lines are the results
using the approximate expression (\ref{goap}) changing $R_{i}$ to $0.138\
r_0 $ and $R_{t}=-R_{i}$. Parameters are $W=7 \hbar^{2}/(2m_e^{*}r_0)$, $%
V=0.48$ eV, $\delta_b=0.032$ eV, $E_{d}^{eff}=0.01$ eV and $q=0.02$. }
\label{fig8}
\end{figure}

In Fig \ref{fig8} we show $\Delta \rho _{f\sigma }\sim \Delta dI/dV$ for an
impurity placed at a distance $0.146\ r_{0}$ to the center (corresponding to
a maximum of the conduction density of states, see Table \ref{table1}), for $%
W=7\hbar ^{2}/(2m_{e}^{*}r_{0})$, and other parameters similar to Figs. \ref
{fig3}(b) or \ref{fig7}. As in previous cases, an intense resonance is
observed not only at $R_{i}$ but also near $-R_{i}$. The maximum of the
absolute value of $\Delta \rho _{f\sigma }$ for the mirage is slightly
displaced towards $0.18r_{0}$.We also show in Fig \ref{fig8} the results
using the approximation for the unperturbed conduction electron Green's
function $G_{s}^{0}$ based on the wave functions for the hard wall, Eq. (\ref
{goap}). The resonance is broader in this case. This is due to the fact that
with decreasing $W$, not only the maximum in real space of the conduction
density of states shifts, but also the spectral weight of the resonance at $%
\epsilon _{F}$ near this maximum decreases, leading to a smaller effective
hybridization. When $W$ is increased so that $\delta _{F}\sim 0.02\ $eV, the
intensity at $-R_{i}$ increases. Also when $\delta _{b}=0$ is taken,
increasing $V_{s}$ the resonances broaden by $\sim 20\%$. These changes are
qualitatively similar to those reported for the smaller ellipse.

\begin{figure}[tbp]
\centering
\includegraphics[scale=0.5,clip,keepaspectratio]{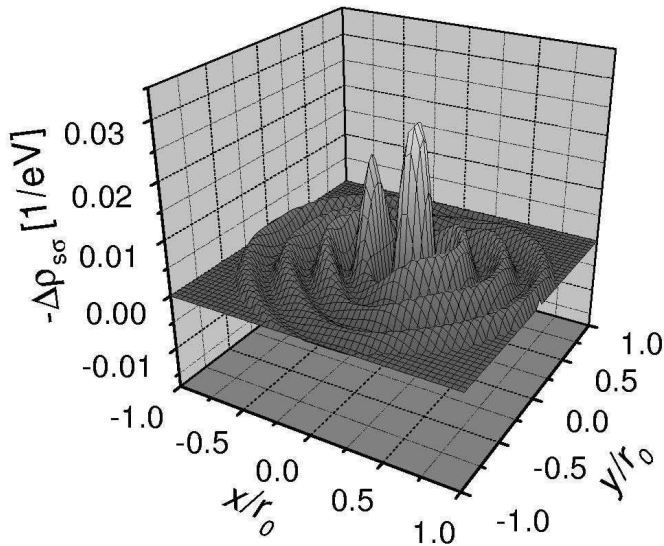}
\caption{ Conduction electron density of states at $\epsilon_{F}$ as a
function of position. Other parameters as in Fig. \ref{fig8}. }
\label{fig9}
\end{figure}

The space dependence of the conduction density of states is shown in Fig. 
\ref{fig9}. The two clear minima near $\pm R_{i}$ can be seen, one of them
corresponding to the impurity and the other to the mirage, in agreement with
experiments \cite{man2}. The addition of the impurity breaks the degeneracy
of the states with $m=\pm 1$ at $\epsilon _{F}$. If the zero line of the
angular variable $\theta $ is aligned with the impurity, the wave function
which hybridizes with the impurity has an angular dependence proportional to 
$\cos \theta $, and the remaining one, proportional to $\sin \theta $ does
not hybridize. This has led to the proposal \cite{man2} that two independent
and simultaneous mirages can be observed placing two impurities at the same
distance from the center and forming an angle of $90$ degrees. While this
seems in agreement with experiment, the conduction states with even $m$
(which lie out of $\epsilon _{F}$), introduce a small interaction between
both channels.

\begin{figure}[tbp]
\centering
\includegraphics[scale=0.5,clip,keepaspectratio]{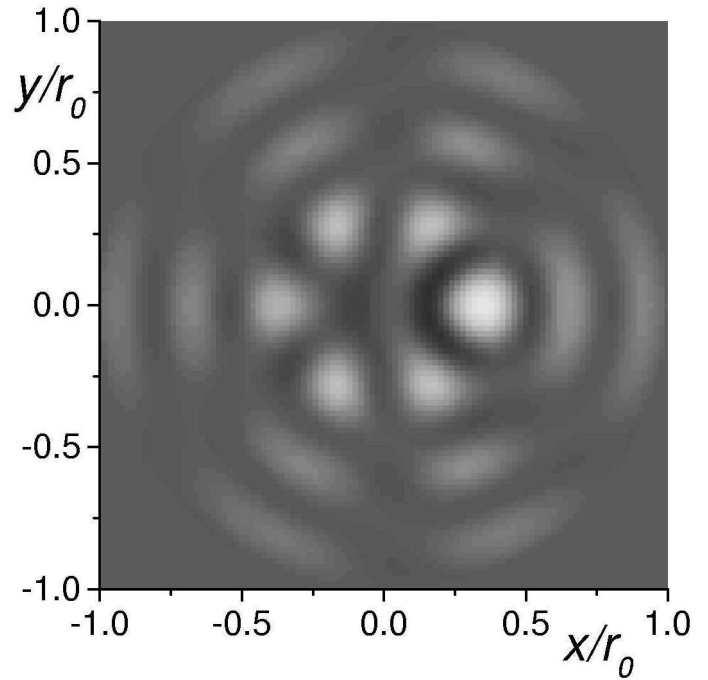}
\caption{ Contour plot of the conduction electron density of states at $%
\epsilon_{F}$ as a function of position for a smaller corral (see text).
Parameters are $W=15 \hbar^{2}/(2m_e^{*}r_0)$, $V=0.45$ eV, $\delta=0.035$
eV, and $E_{d}^{eff}=0.013$ eV.}
\label{fig10}
\end{figure}

To end this section we discuss what happens when $r_{0}$ is slightly reduced
in such a way that the degenerate $35^{th}$ and $36^{th}$ conduction
eigenstates lie at the Fermi energy. These states have $m=\pm 3$ so that one
expects five different mirages as a function of the angle $\theta $, located
near $r=0.32r_{0}$, where the eigenstates for a hard wall corral are peaked.
Since the value of the wave function at these extrema is nearly 3/4 times
smaller than in the previous case, the effective hybridization of the
impurity with the states at the Fermi energy is reduced. As a consequence,
the Kondo depression observed in the energy dependence of $\Delta dI/dV$ is
narrower (with a total width of $\sim $ 0.008 eV for $\delta _{F}\sim 0.02\ $%
eV and $\delta _{b}\neq 0$), and the intensity at the mirages is smaller. In
Fig. \ref{fig10} we show the space dependence of $\Delta \rho _{s\sigma
}\sim \Delta dI/dV$ for parameters near those of Fig. 4. A potential $W$
corresponding to harder walls (leading to $\delta _{F}\sim 0.02\ $eV) than
in Fig. 9 has been chosen in order that the different mirages can be clearly
seen. The most intense ones are those for $\theta =\pm 120$ degrees. In
addition to the five depressions in $\Delta \rho _{s\sigma }(R_{t})$
corresponding to the mirages, one can see a ring of {\em positive} $\Delta
\rho _{s\sigma }$ centered at the impurity site at $R_{i}=0.337r_{0}$ and
radius smaller than $r_{0}$. The intensity of this ring increases with
decreasing $W$ (softer walls) and is therefore also apparent in Fig. \ref{fig9}.
We believe that this feature is probably related with the Friedel
oscillations which are observed experimentally when the impurity is placed
on the clean surface \cite{knorr}.
\section{Spin-spin correlations}

The conventional view of the Kondo effect for an impurity in an ordinary
metal, interprets it in terms of a magnetic screening cloud around the
impurity of radius

\begin{equation}
\xi =\hbar v_{F}/T_{K},
\end{equation}
where $v_{F}$ is the Fermi velocity. The existence of this cloud is still
controversial \cite{sor,barz,col}. Recent theoretical work has shown that
the persistent current as a function of flux $j(\Phi )$ in mesoscopic rings
with quantum dots changes its shape smoothly as the length of the ring $L$
goes through $\xi $ and that $jL$ is a universal function of $L/\xi $ \cite
{pc,aff}.

In spite of this controversy, it is interesting to study the effect of the
corral on the space dependence of spin-spin correlations. To this end we
consider the correlations between the impurity spin ${\bf S}$ and that of
the surface conduction states at position $r$, ${\bf s}(r)$

\begin{equation}
\langle {\bf S\cdot s}(r)\rangle =\sum_{j\alpha \beta \gamma \delta }\frac{%
\lambda ^{2}}{4}\overline{\varphi }_{j}(r)\varphi _{j}(r)\langle d_{\alpha
}^{\dagger }{\bf \sigma }_{\alpha \beta }d_{\beta }\cdot s_{j\gamma
}^{\dagger }{\bf \sigma }_{\gamma \delta }s_{j\delta }\rangle .  \label{ss1}
\end{equation}
For infinite confinement ($W\rightarrow +\infty $, $\delta _{F}=0$), exact
numerical results in a reduced basis set show that the space dependence of
this function follows closely the probability of the wave function lying at
the Fermi level \cite{wil}. However, as discussed before, this situation is
not realistic.

\begin{figure}[tbp]
\centering
\includegraphics[scale=0.5,clip,keepaspectratio]{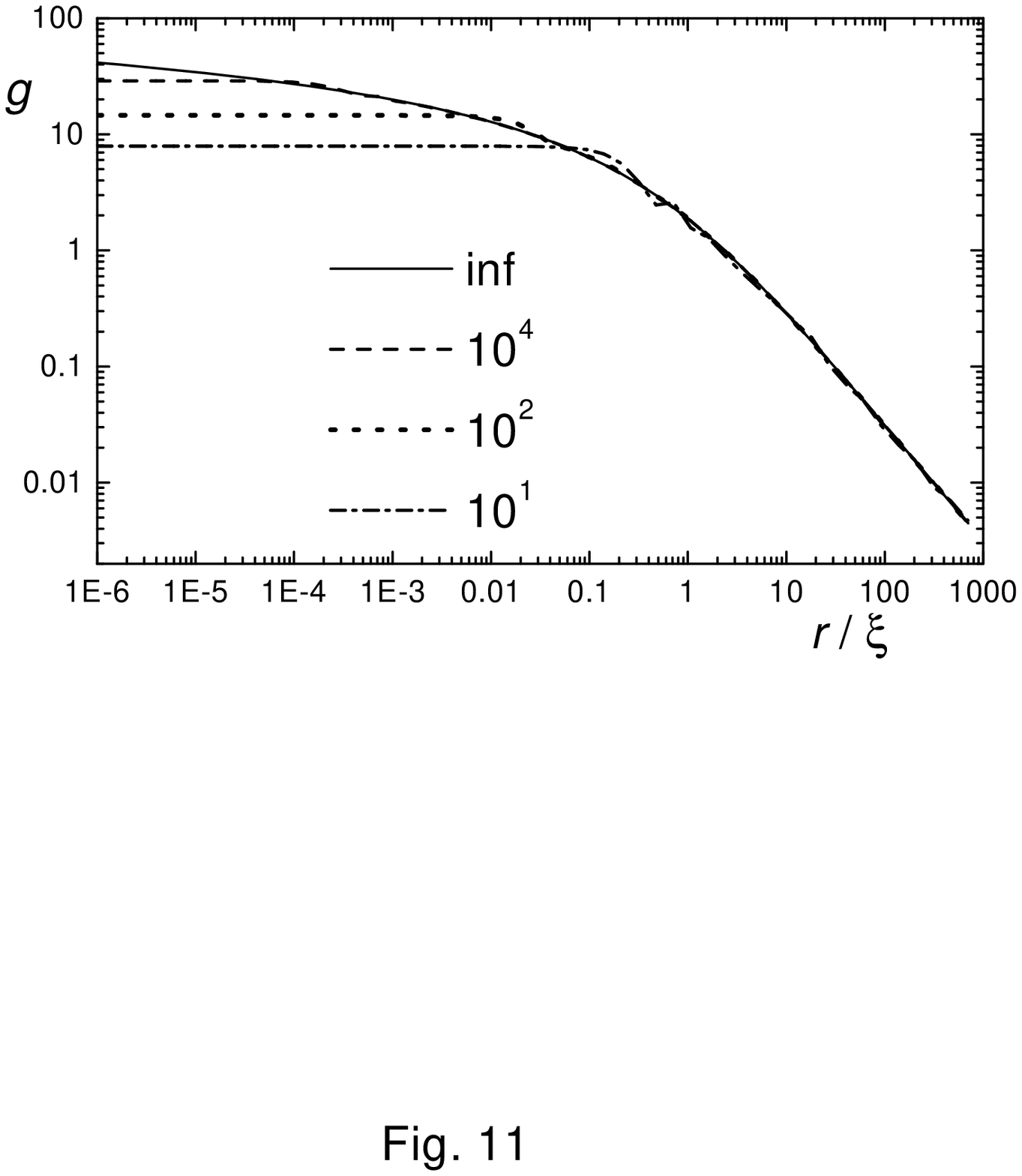}
\caption{Function $g(r/\xi )$ (see Eq. (\ref{g})) for different symmetric
limits of integration indicated inside the figure.}
\label{fig11}
\end{figure}

In the uncorrelated case $U=0$, using Wick's theorem and the symmetry of the
ground state one has:

\begin{equation}
\langle {\bf S\cdot s}(r)\rangle =-\frac{3}{2} \lambda \left| \sum_{j}\varphi
_{j}(r)\langle d_{\sigma }^{\dagger }s_{j\sigma }\rangle \right|^{2}.  \label{sap}
\end{equation}
The saddle point approximation of the slave boson treatment of the Anderson
model \cite{colb}, or of the $1/N$ expansion of the Kondo model \cite{col},
lead to a similar expression. Here we use Eq. (\ref{sap}) as an
approximation for the finite $U$ problem. This amounts to an infinite
partial summation of diagrams which can be separated into two pieces. The
expectation value entering Eq. (\ref{sap}) can be related to $G_{d}(\omega )$
without further approximations using equations of motion. At zero
temperature this leads to:

\begin{equation}
\langle {\bf S\cdot s}(r)\rangle \cong -\frac{3\lambda ^{4}}{2\pi ^{2}}%
\left|\sum_{j}\overline{\varphi }_{j}(R_{i})\varphi _{j}(r)\int^{\epsilon
_{F}}d\omega 
\text{\rm Im}%
\frac{G_{d}(\omega )}{\omega +i\eta -\epsilon _{j}}\right|^{2}  \label{ss}
\end{equation}
We believe that the main qualitative features of the problem are retained by
this approximation, at least for $r\lesssim \xi $ which is the relevant case
for the corrals. To check this, we evaluated this expression for the case of
a Kondo impurity in an ordinary three-dimensional metal, assuming (as usual
in this case) that plane waves describe the conduction electrons, and
(except for a multiplicative constant) $G_{d}(\omega )\sim 1/(\omega
-\epsilon _{F}+iT_{K})$ \cite{por,aga}, with $T_{K}\ll \epsilon _{F}$.
Calling $I(\epsilon _{j})$ the integral entering Eq. (\ref{ss}) and
linearizing the spectrum around $\epsilon _{F}=\hbar v_{F}k_{F}$ one has:

\begin{eqnarray}
\sum_{j}\overline{\varphi }_{j}({\bf R}_{i})\varphi _{j}({\bf r})I(\epsilon
_{j})&\sim& \int d^{3}{\bf k}\exp (i{\bf k}\cdot {\bf r})I(\hbar v_{F}k)\nonumber\\
&\sim&\frac{k_{F}}{r}\int dk\sin (kr)I(\hbar v_{F}k).\nonumber\\
\label{3d}
\end{eqnarray}
Here the vectors are denoted by boldface (${\bf k},{\bf r}$) to distinguish
them from their absolute values ($k,r$). Changing variable $z=(k-$ $%
k_{F})\xi $, and performing the integral in $\omega $ ($I(\hbar v_{F}k)$),
we obtain except for a constant:

\begin{equation}
\langle {\bf S\cdot s}(r)\rangle \sim \frac{\sin ^{2}(k_{F}r)}{r^{2}}g(r/\xi
),  \label{ss3d}
\end{equation}

\begin{equation}
g(r/\xi )=\int dz\cos (\frac{zr}{\xi })\frac{\pi |z|/2-\ln (z)}{z^{2}+1}.
\label{g}
\end{equation}
$g(0)$ diverges unless one introduces a cutoff at $|z|\cong k_{F}\xi
=\epsilon _{F}/T_{K}$. For other values of the argument the integral can be
extended to infinity at both sides, and it can be related to tabulated
integrals \cite{grad}. The result is:

\begin{equation}
g(x)=-\pi e^{x}%
\mathop{\rm Ei}%
(-x)\text{; }x\neq 0,  \label{gei}
\end{equation}
where $%
\mathop{\rm Ei}%
(x)=\int_{-\infty }^{x}dt\exp (t)/t$ is the integral exponential function.
The function $g(r/\xi )$ is represented in Fig.\ref{fig11} for several values of the
cutoff in 
\mbox{$\vert$}%
$z$%
\mbox{$\vert$}%
. As seen in the figure, this cutoff only modifies the function for small
arguments. There is a crossover for $r\sim \xi $ and for $r\gg \xi $, $%
g(r/\xi )$ decays linearly with $r$, leading to a $1/r^{3}$ decay of $%
\langle {\bf S\cdot s}(r)\rangle $. The functional form of $\langle {\bf %
S\cdot s}(r)\rangle $ is very similar to that of the spin susceptibility
suggested by Sorensen and Affleck \cite{sor,barz} for $r\lesssim \xi $.
However, for $r\gg \xi $ they obtain an exponential decay. Therefore, we
expect that the main features of the space dependence of the correlation
function are retained by the approximation Eq. (\ref{sap}) for $r\lesssim
\xi $.
\begin{figure}[tbp]
\centering
\includegraphics[scale=0.4,clip,keepaspectratio]{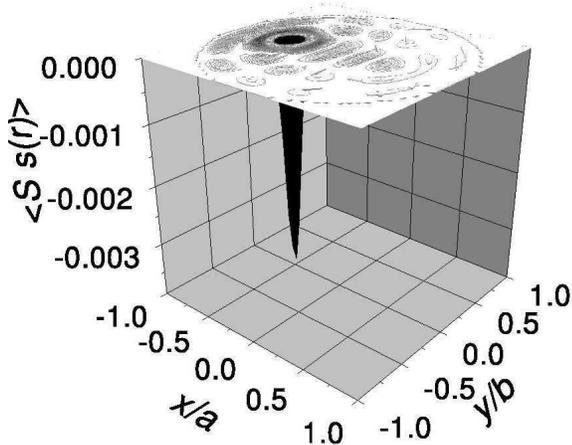}
\caption{ Correlations between the spin of the impurity and that of the
conduction electrons as a function of the position of the latter. Other
parameters as in Fig. \ref{fig4}. }
\label{fig12}
\end{figure}

In Fig. \ref{fig12}, we show the evaluation of Eq. (\ref{ss}) for the elliptical
corral, with the parameters of Fig.\ref{fig4} and zero temperature. In contrast to $%
dI/dV$ or the local spectral density of states near $\epsilon _{F}$ (Fig. \ref{fig4}
and previous calculations for perfect confinement ($\delta _{F}=0$) \cite{wil},
the spin correlations do not recover substantially at the empty
focus, although still faint features related with the conduction state at
the Fermi level ($j=42$) are still discernible.. The ratio of the spin
correlations at both foci is $\sim 1/170$. Note that from $T_{K}$ and $v_{F}$
for surface electrons in the mirage experiments, one gets $\xi \sim 90$ \AA
, slightly larger than the distance between foci. As it is clear from the
calculation above, for $r\lesssim \xi $ all conduction states (also those at
high energies) contribute to the spin-spin correlations. Therefore, the
destructive interference near the empty focus is larger than in the case of $%
dI/dV$, in which mainly unperturbed conduction states near $\epsilon _{F}$
are sampled. The participation of high energy electrons in the formation of
the singlet ground state has been stressed recently \cite{col}.

\section{Summary and discussion}

We have studied the 2D Green's function for free electrons, subject to a
potential that simulates a continuous circular corral. We disregarded the
hybridization with bulk states at the boundary of the corral. In scattering
theories, an imaginary part is introduced for the phase shift to take into
account this effect in phenomenological way \cite{cro,aga,fie}, and this
leads to loss of intensity at the mirage point when an impurity is
introduced in the corral. In our theory this loss of intensity comes as a
result of quantum interference, as explained in Section IV. Although the
density of states is continuous in space and energy, the energy dependence
of the Green's function at low energies can be written as a sum of discrete
poles, representing resonant states. This property is expected to persist
for more realistic potentials which might include hybridization with bulk
states or individual atoms at the boundary at the corral \cite{cal}. The
introduction of a width $\delta _{i}$ of each resonance is the main
difference with a calculation assuming a hard wall corral. The energies $%
\epsilon _{i}$ and the space dependence of the density of states inside
corral are well described by the calculation assuming hard walls, but the
magnitude of the latter is somewhat overestimated. Our model leads to a
width of each resonance $\delta _{i}$, which is to a first approximation,
proportional to its energy $\epsilon _{i}$ . We must warn however that at
least above the Fermi energy, the surface states have an intrinsic width,
even in absence of the corral \cite{hub}. Thus, while we expect that $\delta
_{i}$ can be reduced for example by depositing a second line of atoms at the
boundary of the corral, there are probably intrinsic limits to this
reduction. Another shortcoming of our one-body calculation is that actually
the effective mass depends on energy. Photoemission experiments suggest a
flattening of the dispersion above $\epsilon _{F}$ \cite{hub}. This is also
apparent in the comparison of experimental energy levels for a circular
corral an those of a hard wall calculation \cite{cro} and persists in our
results for soft boundaries.

The expansion as a sum of discrete poles of the conduction electron Green's
function is a convenient starting point for a perturbative treatment of the
Anderson model that describes the physics when a magnetic impurity is
introduced inside the corral. The line shape of the change in differential
conductance $\Delta dI/dV$ and the relative intensity at points distant from
the impurity (where a ``mirage'' of the impurity is observed \cite{man}) is
very sensitive to $\delta _{i}$. This sensitivity is reduced but persists if
an important intrinsic width $\delta _{b}$ of the impurity level due to
direct hybridization with bulk states is introduced. A smaller $\delta _{i}$
leads to a stronger intensity at the mirage point, and the space dependence
tends to that of the conduction state at $\epsilon _{F}$. However, for very
small $\delta _{i}$ the line shape (voltage dependence), is strongly
distorted and two peaks of positive weight of $\Delta dI/dV$ appears at
moderate non-zero voltages. The line shape and its width depend also on the
particular conditions of the experiment. This fact includes the case of the
clean surface, as observed experimentally \cite{man}, and calculated before 
\cite{ali2}. A larger width and more intense resonance is expected when the
wave function of the conduction state which lies at the Fermi energy has a
stronger amplitude at the impurity.

While the {\em space} dependence of $\Delta dI/dV$ is determined only by the
conduction electron Green's function in absence of the impurity (see Eqs. (%
\ref{didv}) and (\ref{gf})), the {\em voltage} dependence is very sensitive
to the impurity Green's function. Most previous theories either assume or
fit this line shape \cite{por,aga,fie} or are unable to explain the observed
one because $\delta _{i}=0$ was assumed \cite{hal}. Also while numerical
diagonalization with $\delta _{i}=0$ and a reduced basis set show spin-spin
correlations which remind the conduction state at the Fermi level \cite{wil}%
, we obtain using additional approximations discussed in Section V, that no
appreciable projection to the mirage point is present in these correlations.
This is due to the fact that the screening of the spin involves conduction
states far from the Fermi energy $\epsilon _{F}$, while $\Delta dI/dV$
selects states near $\epsilon _{F}$. These experiments seem unable to settle
the still controversial issue of the Kondo cloud \cite{sor,barz,col}.
Instead, a crossover as a function of size is expected in mesoscopic rings 
\cite{pc,aff}.

Our many-body theory is able to explain the main features of the space and
voltage dependence of $\Delta dI/dV$ in several experiments involving the
quantum mirage\cite{man,man2} and predictions for other geometries and
experimental conditions were made. While it would be interesting to test
this predictions and study the effects of temperature, the theory has some
limitations. A real Co atom added to a Cu(111) surface has degenerate
localized 3d levels with strong correlations. We have taken only one d
orbital with a moderate on-site Coulomb repulsion $U=1$ eV to be able to use
the perturbative approach. We believe however that these shortcomings affect
quantitative details but do not invalidate our conclusions. Previous results
using larger values of $U$ and a different technique led to results very
similar to the perturbative ones \cite{wil}. A Co atom also has extended 4s
and 4p electrons, which should have a large hopping to the STM tip when it
is near the impurity. This should affect the amplitude of $\Delta dI/dV$ at
moderate distances to the impurity. This effect has so far been neglected in
the theories of the mirage effect.

\section*{Acknowledgments}

This work was sponsored by PICT 03-06343 of ANPCyT. One of us (AAA) is
partially supported by CONICET.


\begin{references}
\bibitem{note}  Experimentally $\epsilon _{4}^{1}\cong \epsilon _{F}$ for $%
r_{0}=63.5$ \AA\ and finite $W$, while we have taken $\epsilon
_{4}^{1}=\epsilon _{F}$ for $W\rightarrow +\infty $. Then, actually we have
to start with a slightly larger effective mass to reach the experimental
condition. However, since the difference is small, practically it does not
modify our results and we neglect it.
\bibitem{notepol}  To find numerically the poles of $G_{d}^{0}(\omega )$ we
started form the case $\delta _{F}=0$ (for which $N-1$ poles of $%
G_{d}^{0}(\omega )$ lie between two consecutive poles of $G_{s}^{0}(\omega )$%
) and then increased slowly $\delta _{F}$ to the desired value. Doing so, we
can trace continuously the position of these poles in the complex plane.
\bibitem{note2}  In general, the unperturbed Green's function (retarded or
time-ordered) for $d$ electrons can be written as $(G_{d}^{0}(\omega
))^{-1}=\omega -E_{d}^{eff}-\sum_{k}|V_{k}|^{2}G_{k}^{0}(\omega )$, where $k$
labels all one-electron eigenstates in absence of the impurity and the
time-ordered $G_{k}^{0}(\omega )$ = $[\omega -\epsilon _{k}+i\eta $sgn$%
(\epsilon _{k}-\epsilon _{F})]^{-1}$ = $P[1/(\omega -\epsilon _{k})]-i\pi
\delta (\omega -\epsilon _{k})$sgn$(\omega -\epsilon _{F})$. Then, for real $%
\omega $, Im$(G_{d}^{0}(\omega ))^{-1}$ = sgn$(\omega -\epsilon _{F})\pi
\sum_{k}|V_{k}|^{2}\delta (\omega -\epsilon _{k})$. The retarded Green's
functions have the same form removing the sign (sgn) functions.
\end{references}
\end{document}